\documentclass{IEEEtran}

\author{\IEEEauthorblockN{Joseph Daws Jr.\IEEEauthorrefmark{1},
Armenak Petrosyan\IEEEauthorrefmark{2},
Hoang Tran\IEEEauthorrefmark{2}, 
Clayton G.~Webster\IEEEauthorrefmark{1}\IEEEauthorrefmark{2}~} 

\IEEEauthorblockA{\IEEEauthorrefmark{1}Department of Mathematics, University of Tennessee Knoxville,
Knoxville, Tennessee 37996--1320 USA
\IEEEauthorblockA{\IEEEauthorrefmark{2}Computational and Applied Mathematics, 
Oak Ridge National Labratory, Oak Ridge, Tennessee 37831--6164 USA}
} 
} 
\usepackage{amsthm}
\usepackage{amsmath,amssymb}
\usepackage{enumitem}
\usepackage{graphicx}
\usepackage{subcaption}
\usepackage{float}
\usepackage{bm}
\usepackage[normalem]{ulem}
\usepackage{hyperref}
\usepackage{listings}
\usepackage{comment}
\hypersetup{
	colorlinks,
	citecolor=blue,
	filecolor=blue,
	linkcolor=blue,
	urlcolor=blue
}
\newtheorem{theorem}{Theorem}

\newtheorem{proposition}{Proposition}
\newtheorem{lemma}{Lemma}

\newtheorem{definition}{Definition}
\newtheorem{remark}{Remark}

\newcommand{\R}{\mathbb{R}}

\newcommand{\CC}{\mathbb{C}}

\def\conv{{\textnormal conv\,}}
\def\supp{{\text{supp}\,}}

\newcommand{\eps}{\varepsilon}

\begin{document}

\title{A Weighted $\ell_1$-Minimization Approach For Wavelet Reconstruction of Signals and Images}

\maketitle

\begin{abstract}
In this effort we propose a convex optimization approach based on weighted $\ell_1$-regularization for reconstructing objects of interest, such 
as signals or images, that are sparse or compressible in a wavelet basis. 
We recover the wavelet coefficients associated to the 
functional representation of the object of interest by solving our proposed optimization problem.
We give a specific choice of weights and show numerically that the chosen weights 
admit efficient recovery of objects of interest from either a set of sub-samples or
a noisy version. Our method not only exploits sparsity but also 
helps promote a particular kind of structured sparsity often exhibited by many 
signals and images. Furthermore, we illustrate the effectiveness of the proposed convex 
optimization problem by providing numerical examples using both orthonormal wavelets and 
a frame of wavelets. We also provide an adaptive choice of
weights which is a modification of the iteratively reweighted $\ell_1$-minimization 
method introduced in \cite{reweightedl1}.
\end{abstract}

\section{Introduction}
We investigate recovering an object of interest (OoI) from either a small number of samples 
or a noisy version using a weighted $\ell_1$-norm regularized convex optimization scheme
with a specific choice of weights. 
Throughout this effort, the functional representation of an OoI is given by
	\begin{equation} \label{eq:approximation_expansion}
		f(\bm{y}) := \sum_{\bm{\nu} \in \mathcal{S}} c_{\bm{\nu}} \Phi_{\bm{\nu}}(\bm{y}) 
		+ \sum_{\bm{\nu} \in \mathcal{W}} c_{\bm{\nu}} \Psi_{\bm{\nu}}(\bm{y}),
	\end{equation}
where $\bm{y}$ is in the domain $\mathcal{U}$ of $f$, 
$\mathcal{S}$ and $\mathcal{W}$ are two finite sets of multi-indices
which we will specify later, $\{ \Phi_{\bm{\nu}} \}_{\bm{\nu \in \mathcal{S}}}$ 
is a family of scaling functions, $\{ \Psi_{\bm{\nu}} \}_{\bm{\nu \in \mathcal{W}}}$ is a family of wavelet 
functions, and $c_{\bm{\nu}}$ is either a wavelet or scaling function coefficient. 
We will discuss the wavelet and scaling functions in Section \ref{sec:theoretical_discussion}. 
The recovery of $f$ is achieved by identifying a vector of coefficients, 
$\bm{c} :=(c_{\bm{\nu}})_{\bm{\nu} \in \mathcal{S} \cup \mathcal{W}}$, 
from our proposed convex optimization problem. The weighted $\ell_1$-norm, 
$\| \cdot \|_{\bm{\omega},1}$ is defined as 
	\begin{equation}
		\| \bm{c} \|_{\bm{\omega},1} = \sum_{{\bm{\nu}} \in \mathcal{J}} \omega_{\bm{\nu}} |c_{\bm{\nu}}|,
	\end{equation}
given the vector of $N$ weights $\bm{\omega} = (\omega_{\bm{\nu}})_{{\bm{\nu}} \in \mathcal{J}}$
where $\mathcal{J} := \mathcal{S} \cup \mathcal{W}$ and the cardinality of $\mathcal{J}$ is $N$. 
The coefficients $\bm{c}$ are obtained by solving 
	\begin{equation} \label{eq:weighted_l1_min}
		\min_{\bm{c} \in \CC^N} \lambda \| \bm{c} \|_{\bm{\omega},1} 
		+ \| \bm{A} \bm{c} - \tilde{\bm{f}} \|_{2}^2,
	\end{equation}
where $\bm{f} = (f(\bm{y}_1),\dots,f(\bm{y}_m))$ is an $m \le N$-dimensional vector of evaluations 
of $f$ at the points $\bm{y}_i \in \R^d$ which may or may not be noisy, 
$\tilde{\bm{f}}$ is the scaled vector $\tilde{\bm{f}} = \bm{f}/\sqrt{m}$ and 
$\bm{A}$ is the $m \times N$ matrix whose entries are
	\begin{equation} \label{eq:measurement_matrix}
		A_{i, \rho({\bm{\nu}})} = \left \{ 
		    \begin{array}{cc}
		        \frac{\Phi_{\rho(\bm{\nu})}(\bm{y}_i)}{\sqrt{m}} & \text{ if } \bm{\nu} \in \mathcal{S} \\ 
	    	    \frac{\Psi_{\rho(\bm{\nu})}(\bm{y}_i)}{\sqrt{m}} & \text{ if } \bm{\nu} \in \mathcal{W},
	    	\end{array}
	    \right .
	\end{equation}
given the bijective mapping $\rho: \mathcal{J} \rightarrow \{1,\dots,N\}$,
$m$ evaluation points $\{\bm{y}_i\}_{i=1}^m \subset \R^d$ and $\bm{\nu} \in \mathcal{J}$.
The parameter $\lambda$ in \eqref{eq:weighted_l1_min} controls the trade off between 
the regularization of the solution enforced by the weighted $\ell_1$-norm and the
fidelity to the observation $\bm{f}$ enforced by the $\ell_2$-norm.

\begin{figure} 
    \centering
    \begin{subfigure}[b]{0.45\textwidth}
        \includegraphics[width=\textwidth]{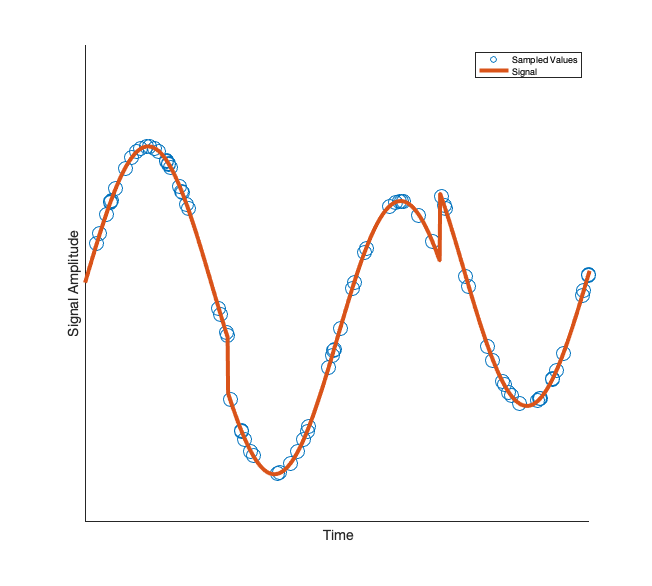}
        \caption{}
        \label{fig:sine_and_subsam}
    \end{subfigure}
    ~ 
    \begin{subfigure}[b]{0.45\textwidth}
        \includegraphics[width=\textwidth]{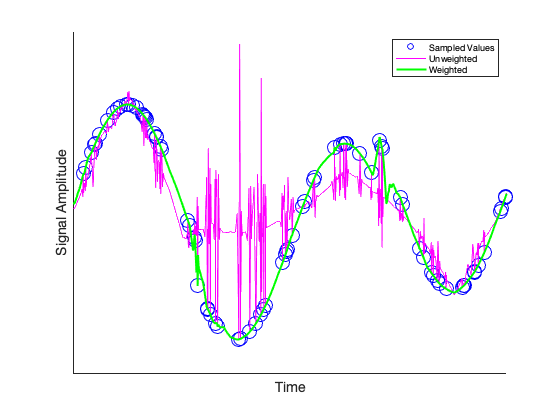}
        \caption{}
        \label{fig:recovs}
    \end{subfigure}
    \caption{Reconstruction of the original signal with both weighted and unweighted $\ell_1$-minimization.
    Here we plot: in Figure (\ref{fig:sine_and_subsam}) the piecewise smooth signal 
    where the circles indicate 80 randomly subsampled values; and in Figure
    (\ref{fig:recovs}) the reconstruction from the 80 subsampled values using weighted and unweighted $\ell_1$-minimization.}
    \label{fig:weighted_vs_unweighted}
\end{figure}

\begin{figure}
    \centering
    \begin{subfigure}[b]{0.45\textwidth}
        \includegraphics[width=\textwidth]{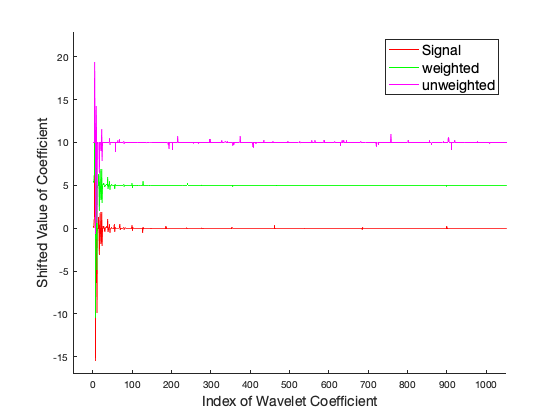}
        \caption{}
        \label{fig:all_coefs}
    \end{subfigure}
    ~ %
    \begin{subfigure}[b]{0.45\textwidth}
        \includegraphics[width=\textwidth]{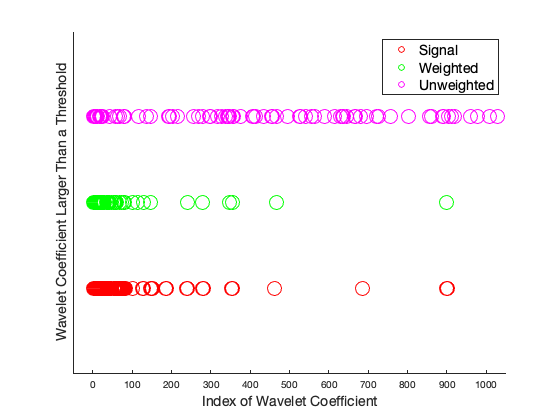}
        \caption{}
        \label{fig:thresh_coefs}
    \end{subfigure}
    \caption{A visualization of how weighted $\ell_1$-minimization recovers a set of coefficients whose sparsity is structured similarly to the original signal.
    The coefficients plotted here are associated wtih the Daubechies $3$ wavelet basis also denoted as as $db3$. 
    For a construction of this wavelet see \cite{book:10_lec_daubechies}. 
    Here we plot: in Figure (\ref{fig:all_coefs}) the values of all wavelet coefficients where the coefficients recovered by 
    unweighted and weighted $\ell_1$-minimization
    are shifted so that their differences are more readily seen; and in Figure (\ref{fig:thresh_coefs}) 
    the coefficients whose magnitudes are larger than $0.01$.}
    \label{fig:weighted_vs_unweighted_coefs}
\end{figure}

The effectiveness of $\ell_1$-minimization is highlighted by its use in compressed sensing 
(CS) \cite{original_cs_paper,donoho_compressed_sensing} and has been successfully deployed 
in many applications such as photography \cite{single_pix_camera}, 
medical imaging \cite{cs_mri} or radar and electromagnetic imaging \cite{cs_electromagnetics}. 
Wavelet representations are extensively employed in data compression and denoising
\cite{wavelet_shrinkage,wavelet_compress_and_denoise}. Despite these triumphs, 
standard, unweighted $\ell_1$-minimization, i.e., the minimization problem \eqref{eq:weighted_l1_min}
where $\bm{\omega} = (1,\dots,1)$, does not seem suitable for the recovery of wavelet coefficients
even for functions with sparse or compressible representations in a wavelet basis. Consider Figure 
\ref{fig:sine_and_subsam} where a piecewise smooth function is plotted. 
As seen in Figure \ref{fig:all_coefs}, many of its coefficients 
are relatively small (only $95$ out of the
$1053$ plotted coefficients have magnitude larger than $0.01$), so this function is compressible in wavelet basis. 
The indices of these large coefficients are given in 
Figure \ref{fig:thresh_coefs}. 
From Figure \ref{fig:recovs} which plots the recovery of the piecewise smooth function from $80$ randomly chosen
samples, it is readily seen that using unweighted $\ell_1$-minimization is not satisfactory. 
Comparing the distribution of the large wavelet coefficients 
recovered by unweighted $\ell_1$-minimization to those of 
the original signal, shown in Figure \ref{fig:all_coefs}, 
it is clear that the unweighted approach leads to the 
recovery of spurious large coefficients that do not correspond to the true signal's coefficients. 
Figure \ref{fig:thresh_coefs} shows the indices of the $123$ coefficients larger 
than the threshold $0.01$ recovered by unweighted $\ell_1$-minimization. 
In particular, we notice that most of the large coefficients of the original signal
are those with low indices, whereas the large coefficients recovered by 
unweighted $\ell_1$-minimization are more uniformly distributed. 

In this effort, we study a model for the structured sparsity of wavelet coefficients of OoI's 
and consider several choices of weights chosen in a particular way which encourage that structure. 
We will use the weights
    \begin{equation} \label{eq:choice_of_weight_alpha}
        \omega_{\pmb{\nu}} = \left \{ 
            \begin{array}{cc}
                \| \Phi_{\bm{\nu}} \|_{L_\infty} & \text{ if } \bm{\nu} \in \mathcal{S} \\ 
                \| \Psi_{\bm{\nu}} \|_{L_\infty} & \text{ if } \bm{\nu} \in \mathcal{W}
            \end{array}
        \right . .
    \end{equation}
This choice is inspired by \cite{high_d_poly_approx} where
recovering the polynomial coefficients of high-dimensional functions by weighted $\ell_1$-minimization
is considered, and the indices of large polynomial coefficients of smooth functions typically
fall in certain kinds of sets called ``lower sets". They show that using  
\eqref{eq:choice_of_weight_alpha} vastly improves the recovery of the functions by
proving that the recovered vector of coefficients has support which is very close to 
a lower set. In other words, the choice of weights promotes 
structure in the recovered coefficients.
The same choice of weights, but definied with respect to wavelet functions in stead of polynomial ones, 
also promotes structure of wavelet coeffienits. Consider Figure \ref{fig:thresh_coefs} which
compares the indices of the $66$ coefficients larger than
the threshold $0.01$ for the original signal, those recovered
by unweighted $\ell_1$-minimization, and those recovered by weighted $\ell_1$-minimization. 
Notice that the distribution of those coefficients recovered by weighted $\ell_1$-minimization 
more closely resembles the distribution of the coefficients of the original signal. 
Furthermore, this choice of weights makes weighted $\ell_1$-minimization robust in the sense that
the recovered sparse vector is close to the true coefficients even when the measurements have been
perturbed by noise. Our numerical examples in Section \ref{sec:numerics} show that weighted 
$\ell_1$-minimization improves recovery for both inpainting and denoising, and
encourages structured sparsity associated with wavelet coefficients. We also consider
solving the inpainting problem using a frame of wavelets. 

In this effort we also provide a choice of weights which can adapt 
to the structure of the wavelet coefficients of a given OoI.
Since wavelets functions are scaled, shifted versions of a mother wavelet,
the weights \eqref{eq:choice_of_weight_alpha} depend only the scale of the 
associated coefficient. More complicated structures beyond the parent-child relationship
may exist. That is, coefficients with large values are not randomly distributed within each scale. 
They may depend on other values within the same scale in addition to those
on adjacent scales. Intuitively, improved performance can be obtained by choosing weights which are 
adapted to the inherent structure of a given set of wavelet coefficients both across and within scale.
We consider a modification of iterative reweighted $\ell_1$-minimization (IRW $\ell_1$-minimization), 
introduced in \cite{reweightedl1}, where a sequence of weighted $\ell_1$-minimization problems are solved. 
The weights used in IRW $\ell_1$-minimization are updated based on the previously recovered vector of coefficients. 
Our modification to IRW $\ell_1$-minimization described in Section \ref{sec:theoretical_discussion}
updates the weights based on both the scale of the associated coefficients and the value of the
coefficients recovered at the previous iteration. 
Our numerical examples which follow show that this adaptive choice of weights 
produces better results at the cost of solving several weighted $\ell_1$-minimization
problems. 

\subsection{Related Results}
Compressed Sensing based approaches for recovering a function from a limited
collection of measurements or evaluations of a function were considered in
\cite{model_based_cs,bui2015, Candes2008, dwb2008, LI2017, polania2014,fourcartCSbook}
among others. Many of these works use the underlying assumption that the OoI can be
well approximated by an expansion like \eqref{eq:approximation_expansion} were only a few
coefficients are large. Both the recovery of signals using weighted $\ell_1$ minimization 
and the use of structured sparsity have also been considered previously.
For example, \cite{WardWeighted} studies a weighted $\ell_1$ approach and proposes some conditions for the weights, 
but does not provide a specific choice. An iterative process for choosing adaptive weights 
was introduced in \cite{reweightedl1} where weights are updated
based on the coefficients recovered on the previous iteration. 
A specific choice of weights is given in \cite{high_d_poly_approx}
which yields a quantifiable improvement to the sample 
complexity. Binary weights are considered in \cite{cs_electromagnetics}.
A general class of structured sparse signals is considered in \cite{model_based_cs}, 
where the authors establish a recovery guarantee with complexity estimates for two kinds 
of greedy algorithms. Another example where the structure of the wavelet trees is utilized is 
\cite{bui2015}, where a novel, Gram-Schmidt process inspired implementation of an orthogonal matching 
pursuit algorithm is developed. 
The practicality of using sparse tree structures for real world signals has also 
been shown. The work \cite{polania2014} uses Compressed Sensing based recovery of the wavelet 
coefficients of electrocardiogram signals.
Under certain structured sparsity assumption on the representation coefficients 
the authors in \cite{adcock2018oracletype,ADCOCK_2017} show that optimal sampling complexity can be 
achieved by unweighted $\ell_1$-minimization if special sampling strategy is adopted. 
In particular this applies to the inpainting problem, 
however, in our case we assume that the samples are uniform and we do not have the freedom to choose the
sampling strategy.
Moreover, our structured assumption does not fit into their paradigm.

Exploiting the structure of wavelet coefficients has also been used to solve the denoising problem.
Notice that noise added to the measurement $\bm{f}$ principally contributes to the high frequency wavelet
coefficients. Therefore, a naive wavelet denoising scheme is to take the wavelet transform of the noisy
vector $\bm{f}$, threshold the wavelet coefficients and transform back into the original domain. By
thresholding the wavelet coefficients we have removed some high frequency information from the 
wavelet coefficients and therefore we can expect that the some of the noise is also removed. 
More sophisticated thresholding methods have been considered, see e.g., 
\cite{wavelet_shrinkage, Donoho_denoise, denoise_across_scale, bayesian_denoise}.
Whereas these works employ statistical estimation to find important wavelet coefficients, 
our work finds out that with a simple choice of weights which is independent of the OoI, 
we can obtain satisfactory denoising results.
Our proposed weighted $\ell_1$-minimization recovers a vector of coefficients which, 
due to our choice of weights, is less likely to be affected by the high-frequency perturbations 
in the function samples. 

\subsection{Organization}
In Section \ref{sec:theoretical_discussion} we present our choices of weights and
review the relevant research which influenced our approach. We also introduce
a model for wavelet coefficients which futher supports our choice of weights.  

In Section \ref{sec:numerics}, we present some numerical experiments 
which show that an OoI can be successfully recovered using \eqref{eq:weighted_l1_min} 
our specific choices of weights \eqref{eq:choice_of_weight_alpha} and \eqref{eq:wirwl1}.
In particular, we consider the recovery of signals, images, and hyperspectral
images from a set of incomplete measurements. We also solve the denoising problem
for signals and images. 

In Section \ref{sec:conclusion} we discuss possible extensions of this work.

\section{Theoretical Discussion} \label{sec:theoretical_discussion}

\begin{figure} 
    \centering
    \begin{subfigure}[b]{0.47\textwidth}
        \includegraphics[width=\textwidth]{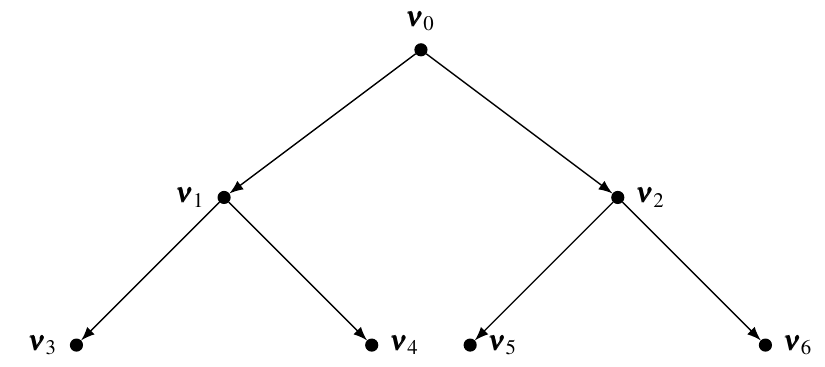}
        \caption{}
        \label{fig:full_tree}
    \end{subfigure}
    ~ 
    \begin{subfigure}[b]{0.47\textwidth}
        \includegraphics[width=\textwidth]{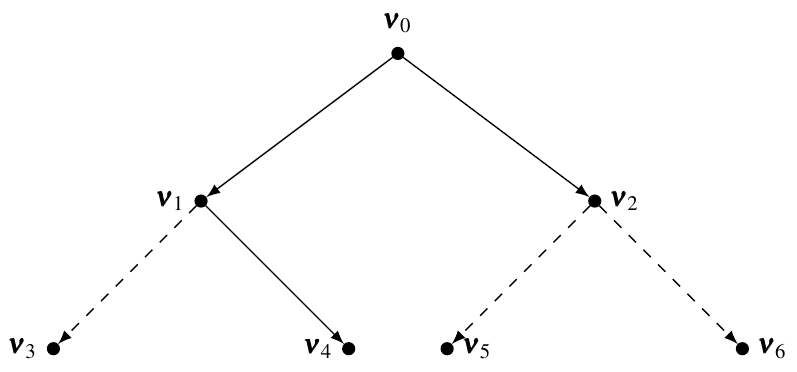}
        \caption{}
        \label{fig:closed_tree}
    \end{subfigure}
    \caption{The wavelet coefficients of real-world signals are associated with 
    $2$-trees. Here we plot: in Figure (\ref{fig:full_tree}) an example of a $2$-tree; and in Figure (\ref{fig:closed_tree}) 
    an example of a subset of nodes of the $2$-tree which forms a closed tree.}
    \label{fig:full_vs_closed}
\end{figure}

In this section we discuss several theoretical elements, which inspired our
choice of weights, that we claim to promote the natural structure exhibited by 
the important wavelet coefficients of real-world OoI. 
Before justifying this claim and presenting a model for wavelet coefficients, 
we will first define $k$-ary trees, which are a special case of a kind of graph called a tree. 
A directed graph is called a tree if it satisfies the following two conditions:
(i) there is a single node, $\bm{\nu}_0$, which is called the root; and,
(ii) there exists one and only one path from $\bm{\nu}_0$ to any other node
$\bm{\nu}$ in the graph \cite{arborescence_def}. 
The indices of the wavelet coefficients can be identified
with a node on a \textit{full $k$-ary tree}, i.e., a tree so that every node has either 
$k$ edges or zero edges leaving it. For example, Figure \ref{fig:full_tree} 
shows an example of a $2$-tree with the indices 
$\{ \bm{\nu_0},\dots,\bm{\nu_6} \}$. In our model, the edges between nodes are directed
and the direction determines a parent-child relationship between nodes. 
We say that node $\bm{\nu}_i$ is the \textit{parent} of node $\bm{\nu}_j$, 
or equivalently, the node $\bm{\nu}_j$ is the \textit{child} of node $\bm{\nu}_i$  
if one of the edges emanating from $\bm{\nu}_i$ terminates at node $\bm{\nu}_j$.
In general, we denote the parent of node $\bm{\nu}_j$ as $p(\bm{\nu}_j)$.
To illustrate, consider Figure \ref{fig:full_tree} where $\bm{\nu}_0$ has two child nodes, 
$\bm{\nu}_1$ and $\bm{\nu}_2$, so that $p(\bm{\nu}_2) = p(\bm{\nu}_1) = \bm{\nu}_0$.
We consider the closed tree model for describing the subsets of large coefficients
of signals and images. 

\begin{definition}[Closed Tree]\label{def:closedtree}
     A multi-index set $T$ is called a closed tree if the 
     following two conditions hold:
         \begin{enumerate}
         	\item Each $\bm{\nu} \in T$ may be uniquely identified with a node on a 
	         $k$-ary tree.
	         \item For each node $\bm{\nu} \in T$, 
	         	\begin{equation*}
				\bm{\nu} \in T \implies p(\bm{\nu}) \in T.
			\end{equation*}
	    That is, if a node is in $T$, then so is its parent.
         \end{enumerate}
\end{definition}

An example of a closed tree is given in Figure \ref{fig:closed_tree}. The
motivation for considering closed trees as a model for wavelet coefficients is three-fold. 
    \begin{itemize}
        \item One can construct orthogonal wavelets from a set of the nested approximation spaces
        called multi-resolution analyses that satisfy certain properties, see for example \cite{HernandezWave}.
        The nested relationship between these induces an association between certain wavelet functions on adjacent levels.
        With appropriate indexing of wavelet function, the parent and child relationship of the closed tree 
        corresponds to this association. 
        
        \item The coefficients of a function expressed in an orthonomral wavelet system 
        are given by 
        the inner product of the function with a wavelet function. In practice, this value 
        is approximated using a quadrature rule. This quadrature can be implemented as a
        linear combination 
        of scaling function coefficients at the previous scale 
        \cite{book:10_lec_daubechies}. Calculating coefficients in this
        way clear associates the value of the coefficient associated with a parent to
        the coefficients associated with its child nodes. 
        
        \item The successful application of hidden markov tree models 
        in works such as \cite{hmtModel,dwb2008,choi2000hidden,crouse1998wavelet} 
        in image and signal processing show that it is beneficial to enforce
        correlation between parent nodes and child nodes.
    \end{itemize}

This model makes rigorous a widely known property of wavelet representation of 
signals and images that nodes associated with small wavelet coefficients
are more likely to have small children and nodes associated 
with large wavelet coefficients may have either large or small children.
In light of this it is natural to find a choice of weights which promotes this structure.
Our choice of weights is inspired by \cite{high_d_poly_approx} where it was proven that polynomial coefficients
that are associated with certain kinds of subsets, called lower sets,
can be recovered with weighted $\ell_1$-minimization with weights equal to 
the uniform norms of the tensor product polynomials associated with the coefficients. 
    \begin{definition}[lower set]
        A multi-index set $\mathcal{S} \subset \mathbb{N}_0^d$ is called a lower set if and only if
        \begin{equation*}
            \bm{\nu} \in \mathcal{S} \text{ and } \bm{\mu} \le \bm{\nu} \implies \bm{\mu} \in \mathcal{S},
        \end{equation*}
        where $ \bm{\mu} \le \bm{\nu}$ is interpreted as $\mu_k \le \nu_k$ for each $k=1,\dots,d$.
    \end{definition}
Closed trees have analogous structure to lower sets in the sense that the parent of
every node in the closed tree is also in the closed tree.
Given a family of pre-defined wavelets, such as Haar, Daubechies, etc., 
the weight given in \eqref{eq:choice_of_weight_alpha} is
    \begin{equation} \label{eq:wavenorm}
       \omega_{\bm{\nu}} = \left \{ 
            \begin{array}{cc}
                 \| \Phi_{\bm{\nu}} \|_{L_\infty} =  2^{jd/2}, & \text{ if } \bm{\nu} \in \mathcal{S}  \\
                 \| \Psi_{\bm{\nu}} \|_{L_\infty} =  2^{jd/2}, & \text{ if } \bm{\nu} \in \mathcal{W}
            \end{array}
        \right . 
    \end{equation}
where the multi-index $\bm{\nu} = (j,k_1,\dots,k_{d})$ and $j$ is 
the level on which the coefficients $c_{\bm{\nu}}$ lies.
In this section we established a structured sparsity model for wavelet coefficients 
and related wavelet and tensor product polynomial representations. 
In the next section we consider using weighted $\ell_1$ to recover 
a signal from incomplete or noisy measurements and justify our approach 
using these connections. 

\subsection{Recovery of OoI from incomplete measurements}
The minimum number of measurements $m$ required for the guaranteed recovery of a 
sparse vector is sometimes called the \textit{sampling complexity} in the compressed
sensing literature. For a measurement scheme arising from a bounded, orthonormal system, 
as in \eqref{eq:measurement_matrix}, the number of samples $m$ required for recovery
using unweighted $\ell_1$-minimization depends on the maximum of the uniform 
norms of the orthonormal system \cite{fourcartCSbook}. That is, let 
	\begin{equation} \label{eq:theta_def}
		\Theta := \max_{\bm{\nu}\in \mathcal{J}} \| \Psi_{\bm{\nu}} \|_{\infty},
	\end{equation}
then whenever $m$ satisfies 
	\begin{equation} \label{eq:theta_bound}
		m \ge \Theta^2s \times \text{ log factors }
	\end{equation}
one can recover the \textit{best s-term approximation} to the target function, 
i.e., an approximation formed by superimposing the $s$ functions from the orthonormal 
system corresponding to the $s$ largest coefficients. This condition is sharp or optimal for many sparse 
recovery problems of interest, for example, from Fourier measurements. 
However, for wavelets and high-dimensional polynomials, $\Theta$ can become so large that renders \eqref{eq:theta_bound} 
useless, see \cite{TranWebster18}.  
Motivated by the need of improved algorithms which can exploit the structure of sparse polynomial 
expansions with better recovery guarantee, \cite{high_d_poly_approx} proposes a weighted $\ell_1$ 
approach where the sampling complexity depends on a quantity
$K(s)$ which is strictly smaller than $\Theta^2 s$. More rigorously, they showed that
	\begin{equation} \label{eq:high_d_sampling_complex}
		m \ge K(s) \times \text{log factors},
	\end{equation}
where
	\begin{equation} \label{eq:def_k_omega}
		K(s) := \sup_{S \text{ is a lower set}, |S| \le s } 
		\left \| \sum_{\bm{\nu} \in S} |\Psi_{\bm{\nu}}|^2 \right \|_{L_{\infty}}
	\end{equation}
is sufficient for the recovery of best $s$ term approximations with lower set structures.
Assuming that an OoI has large wavelet coefficients lying on a closed tree, a similar conclusion
about the sampling complexity of weighted $\ell_1$-minimization \eqref{eq:weighted_l1_min} and 
\eqref{eq:choice_of_weight_alpha} can be made. Let us define the analogous quantity to
\eqref{eq:def_k_omega} for wavelets  
	\begin{equation} \label{eq:def_k_tree}
		K_{\mathcal{T}}(s) := \sup_{\substack{T \text{ is closed tree, } |T|\le s}} 
		\left \| \sum_{\bm{\nu} \in T} |\Psi_{\bm{\nu}}|^2 \right \|_{L_{\infty}}.
	\end{equation}
Then it can be shown that the recovery guarantee is
	\begin{equation} \label{eq:tree_sample_complexity}
		m \ge K_{\mathcal{T}}(s) \times \text{ log factors },
	\end{equation}
and that,
    \begin{equation} \label{eq:compare_tree_sample}
        K_{\mathcal{T}(s)} \le \Theta^2 s.
    \end{equation}
so the sufficient condition on sampling
complexity is improved. 

Unlike the polynomial bases considered in \cite{high_d_poly_approx}, 
the guarantee \eqref{eq:tree_sample_complexity} for wavelet bases
is still too demanding. Moreover, it does not reflect the successful recovery from 
\textit{underdetermined} systems, which is the main objective of a compressed sensing 
approach. We postulate that this is due to the limitation 
of our current analysis technique, and plan to address this issue in future work. 
In experiments, some shown in the following sections, we consistently observe that 
weighted $\ell_1$-minimization is able to reconstruct 
signals and images given a small percentage of pixels. 
Therefore, \eqref{eq:tree_sample_complexity} may be very pessimistic. 
More remarkably, the superiority of our proposed weighted $\ell_1$-minimization approach 
over the unweighted approach is clear.
In fact, our numerical examples show that it performs much better not only
for orthonormal systems of wavelets but also for a frame wavelets which we introduce in 
Section \ref{sec:numerics}.

\subsection{Recovery of OoI from noisy measurements}

Suppose that the samples used for the recovery of a function 
using \eqref{eq:weighted_l1_min} are noisy. In particular, we
assume that $\hat{f}(\bm{y}) := f(\bm{y}) + \eta$ where $\eta$ is modeled 
as a Gaussian noise. 
The denoising problem is to recover $f$ given $\hat{\bm{f}} := ( \hat{f}(\bm{y_k}) )_{k=1}^m$. 
This can be solved by using our proposed weighted $\ell_1$-minimization problem to 
recover the true coefficients of $f$. In Section \ref{sec:numerics}, we give numerical
examples of denoising full, noisy signals and images, i.e., $m=N$.
As mentioned in the introduction, a basic denoising approach is to threshold 
the wavelet coefficients of the noisy signal or image. This simple approach
is effective if the noise level is small. For larger noise levels, more advanced thresholding
algorithms have been proposed which adapt to the signal itself, for example, \cite{wavelet_shrinkage}. 
Our proposed weighted $\ell_1$-minimization problem can be related to an iterative 
weighted soft-thresholding approach, where our choice of weights encourages the 
recovered wavelet coefficients to exhibit structure similar to the original signal. 
According to \eqref{eq:wavenorm}, the deeper a wavelet coefficient lies in the
tree, the larger the weight associated with it is, resulting in more aggressive thresholding. 

\subsection{Scale and Wavelet Aware Iteratively Updated Weights}
\begin{figure*} 
    \centering
    \begin{subfigure}[b]{0.45\textwidth}
        \includegraphics[width=\textwidth]{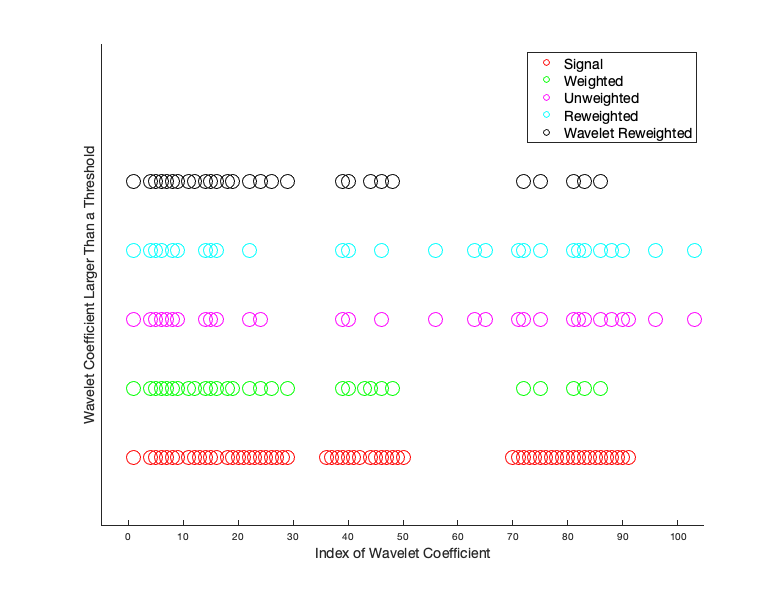}
        \caption{}
        \label{fig:rwl1_thresh}
    \end{subfigure}
    ~ 
    \begin{subfigure}[b]{0.45\textwidth}
        \includegraphics[width=\textwidth]{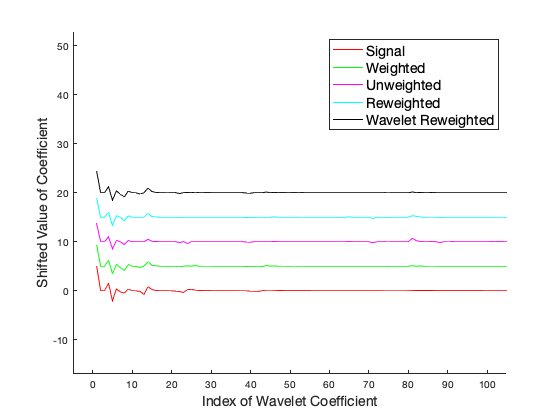}
        \caption{}
        \label{fig:rwl1_coefs}
    \end{subfigure}
    \caption{A comparison of the performance of unweighted, weighted, IRW, and wavelet reweighted 
    $\ell_1$-minimization for recovering the coefficients of a given signal. The IRW example 
    uses the same parameters as \cite{reweightedl1} and the wavelet reweighted example uses the
    weights given in \eqref{eq:wirwl1}.
    } 
    \label{fig:rwl1}
\end{figure*}

Our choice of weights \eqref{eq:choice_of_weight_alpha} naturally 
encourages the property that wavelet coefficients of different scales have
appropriately scaled values. A natural extension would be to pick weights
which take into account the intra-level magnitude correlation of coefficients. 
Although the true wavelet coefficients of an OoI have large and
small values within each scale, our chosen weights do not discriminate between large and small
coefficients within each scale. A method introduced in \cite{reweightedl1} iteratively solves several 
weighted $\ell_1$-minimizations and updates the weights at each iteration based on the recovered
sparse vector, specifically,
    \begin{equation} \label{eq:rwl1}
        \omega_{\bm{\nu}}^{(t)} = \frac{1}{|c_{\bm{\nu}}^{(t-1)}| + \eps} 
    \end{equation}
where $c_{\bm{\nu}}^{(t-1)}$ is the $\bm{\nu}^{th}$ coefficient recovered at step 
$t-1$ and $\eps$ is a parameter that must be chosen. 
Intuitively, this approach tries to find and minimize a concave penalty 
function that more closely resembles $\ell_0$ minimization. 
In practice however, this weighting strategy does
not lead to significantly better results for recovering wavelet coefficients.
In Figure \ref{fig:rwl1}, we see that similarly to the unweighted
$\ell_1$-minimization case, reweighted $\ell_1$-minimization over emphasizes 
coefficients very deep in the wavelet tree leading to poor recovery.
We recreated the results from the paper using the parameters provided by the authors.

\begin{figure}
    \centering
    \includegraphics[width=0.48\textwidth]{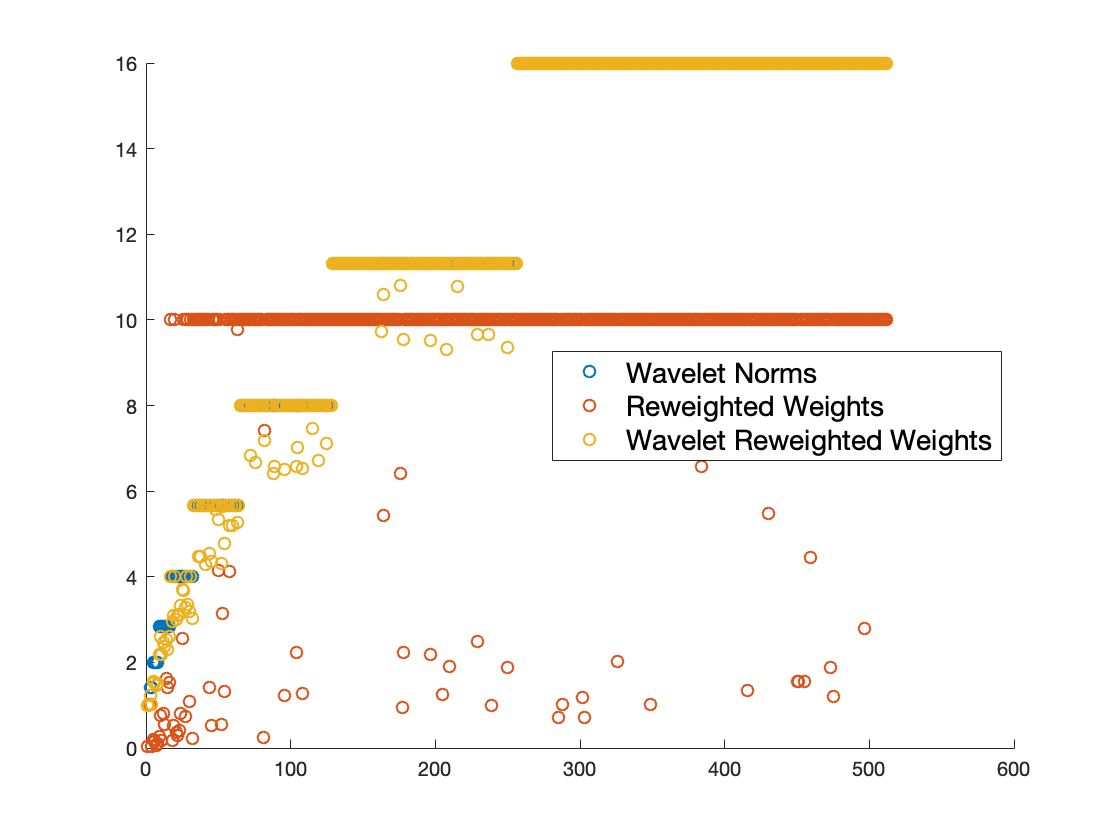}
    \caption{A comparison of the weights used by IRW and wavelet reweighted $\ell_1$-minimization
             after $5$ iterations relative to the choice of weights 
             \eqref{eq:choice_of_weight_alpha}. These weights were obtained in the
             experiment associated with Figures \ref{fig:closed_tree_coef_recov_magnitude} 
             and \ref{fig:closed_tree_coef_recov} descibed in Section \ref{sec:numerics}. }
    \label{fig:compare_weights}
\end{figure}

From \eqref{eq:wavenorm}, it is clear that our choice of weights \eqref{eq:choice_of_weight_alpha}
depend on their level only. On the other hand, notice that the adaptive choice of weights 
used in the usual IRW $\ell_1$-minimization does not take into account the level of the coefficients. 
We propose an alteration of IRW $\ell_1$-minimization, where the weights are updated by the formula
    \begin{equation} \label{eq:wirwl1}
        \omega_{\bm{\nu}}^{(t)} = \omega_{p(\bm{\nu})}^{(0)}
        + 
        \frac{1}{ |c_{\bm{\nu}}^{(t-1)}| + \eps_{\bm{\nu}} },
    \end{equation}
where $\omega_{\bm{\nu}}^{(0)}$ is the weight $\omega_{\bm{\nu}}$ from \eqref{eq:choice_of_weight_alpha}
and $\eps_{\bm{\nu}} := 1 / (\omega_{\bm{\nu}}^{(0)} - \omega_{p(\bm{\nu})}^{(0)})$. 
Observe that the parent of each node is on a shallower level, which implies that 
$\omega_{\bm{\nu}}^{(0)} - \omega_{p(\bm{\nu})}^{(0)} \ge 0$, hence 
$\eps_{\bm{\nu}} \ge 0$. The update \eqref{eq:wirwl1} takes into account both the scale and specific choice of 
wavelet function and can be called scale and wavelet aware iteratively reweighted
$\ell_1$-minimization (hereafter referred to as wavelet reweighted $\ell_1$-minimization).

The motivation for the updates used in wavelet reweighted are twofold. First, on the first iteration,
the weights \eqref{eq:wirwl1} are the same as \eqref{eq:choice_of_weight_alpha}, and therefore, 
they similarly encourage wavelet structured sparsity across levels. On later iterations, 
by \eqref{eq:wirwl1}, $\omega_{\bm{\nu}}^{(t)} \ge \omega_{p(\bm{\nu})}^{(t)}$, hence 
the relative scales of recovered coefficients are maintained.
Second, the term $1/(|c_{\bm{\nu}}^{(t-1)}| + \eps_{\bm{\nu}})$ ensures that 
large coefficients have smaller weights than their sibling coeffients on the 
same scale. Our numerical examples show that the adaptive choice of weights
\eqref{eq:wirwl1} can perform somewhat better than the choice of weights
\eqref{eq:choice_of_weight_alpha}, but at the cost of having to 
solve several weighted $\ell_1$-minimization problems. We also see that it 
consistently performs much better than the usual IRW $\ell_1$-minimization. 

\section{Numerical experiments} \label{sec:numerics}
In this section, we provide numerical results which show the effectiveness of weighted 
$\ell_1$-minimization with our choice of weights 
for the recovery of the wavelet representations of signals, images
and hyperspectral images. 
We also consider the weights
    \begin{equation} \label{eq:choice_of_weight_alpha_pos}
        \omega_{\pmb{\nu}} = \left \{ 
            \begin{array}{cc}
                \| \Phi_{\bm{\nu}} \|_{L_\infty} & \text{ if } \bm{\nu} \in \mathcal{S} \\ 
                \| \Psi_{\bm{\nu}} \|_{L_\infty}^{\alpha} & \text{ if } \bm{\nu} \in \mathcal{W}
            \end{array}
        \right . .
    \end{equation}
Our experiements indicate that choosing $\alpha \ge 1$ consistently performs well, where as 
choosing $0 < \alpha < 1$ consistently performs poorly. There is not much difference
in choosing $\alpha > 1$, therefore the choice $\alpha = 1$ seems to be sufficient in 
general. We additionally present examples related to a frame 
of wavelets for use in the recovery of a signal from partial measurements
as well as experiments using our adaptive
choice of weights \eqref{eq:wirwl1}. 
Recovery of a functional representation of an OoI
\eqref{eq:approximation_expansion} is achieved by identifying the coefficients $\bm{c}$ 
which minimize \eqref{eq:weighted_l1_min}, then applying an inverse discrete 
wavelet transform to $\bm{c}$. The recovered signals and images presented below 
were obtained using \texttt{SPGL1} \cite{spgl1:2007, BergFriedlander:2008} for both the 
unweighted and the weighted cases. The wavelet transforms used are from the 
built-in \texttt{MATLAB} wavelet toolbox.

\subsection{Recovery of synthetic data compressible in wavelet basis}

In this section we consider a synthetic example where the wavelet coefficients of a signal
are exactly supported on a closed tree. We construct such a signal by
randomly choosing a closed wavelet tree with $s$ nodes which is a sub-tree
of a full binary tree with $N=2^J$ nodes. The coefficient values of these $s$ nodes 
are randomly assigned according to a Gaussian distribution whose mean and variance 
depend on the depth on the node. We reconstruct the signal using an 
inverse wavelet transform and randomly sample this signal at $m$ locations. 
These samples are used to recover coefficients using \eqref{eq:weighted_l1_min} 
with several choices of weights, 
IRW $\ell_1$-minimization, and our wavelet reweighted $\ell_1$-minimization. 
Our numerical experiments indicate that 
    \begin{itemize}
        \item the weighted approach outperforms the unweighted approach,
        \item the success of weighted $\ell_1$-minimization
              does not depend too heavily on the choice of $\alpha$, and
        \item our wavelet reweighted approach slightly improves recovery.
    \end{itemize}
    
\begin{figure*} 
    \centering
    \begin{subfigure}[b]{0.45\textwidth}
        \includegraphics[width=\textwidth]{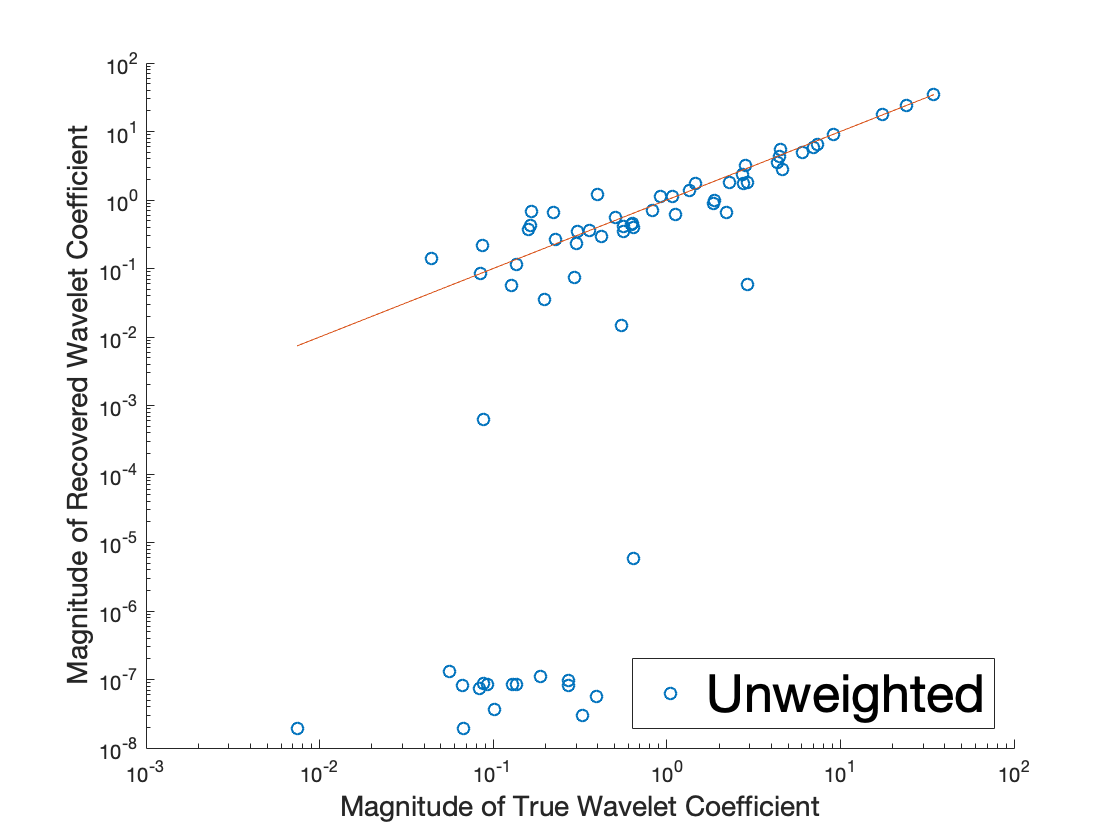}
        \caption{}
        \label{fig:ct_mag_unweighted}
    \end{subfigure}
    ~ 
    \begin{subfigure}[b]{0.45\textwidth}
        \includegraphics[width=\textwidth]{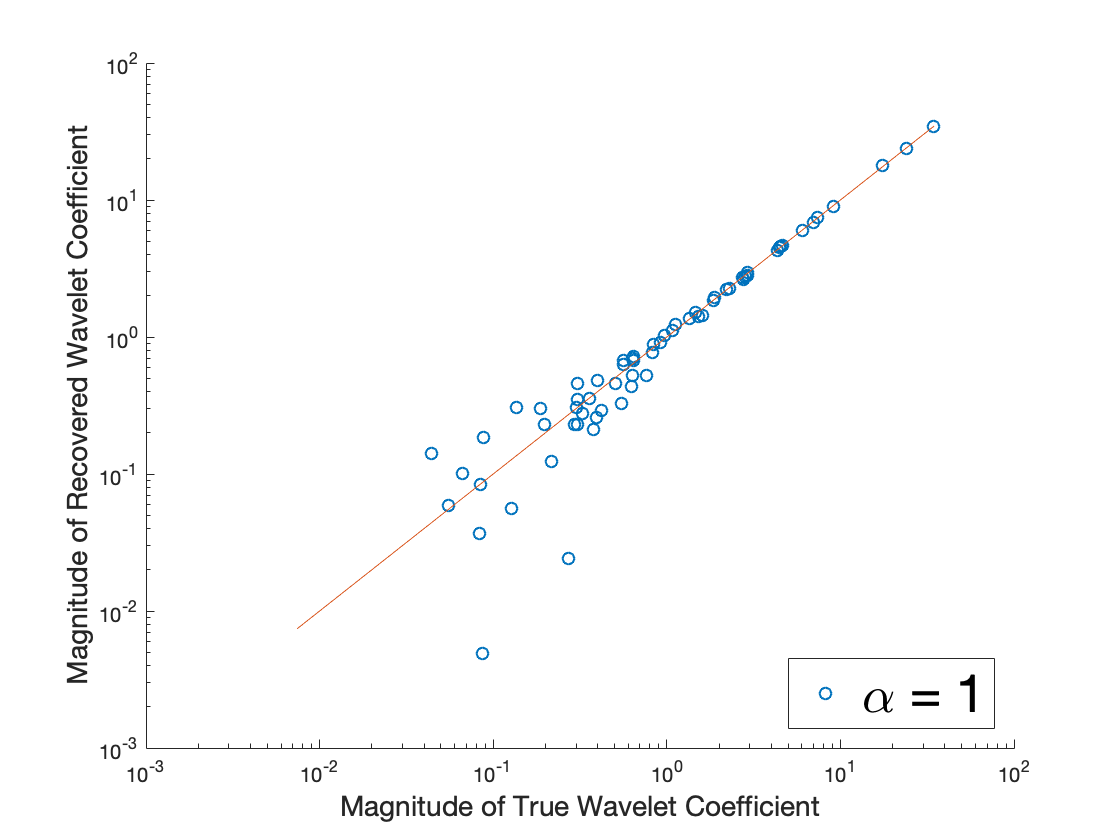}
        \caption{}
        \label{fig:ct_mag_w_half}
    \end{subfigure}
    
    \begin{subfigure}[b]{0.45\textwidth}
        \includegraphics[width=\textwidth]{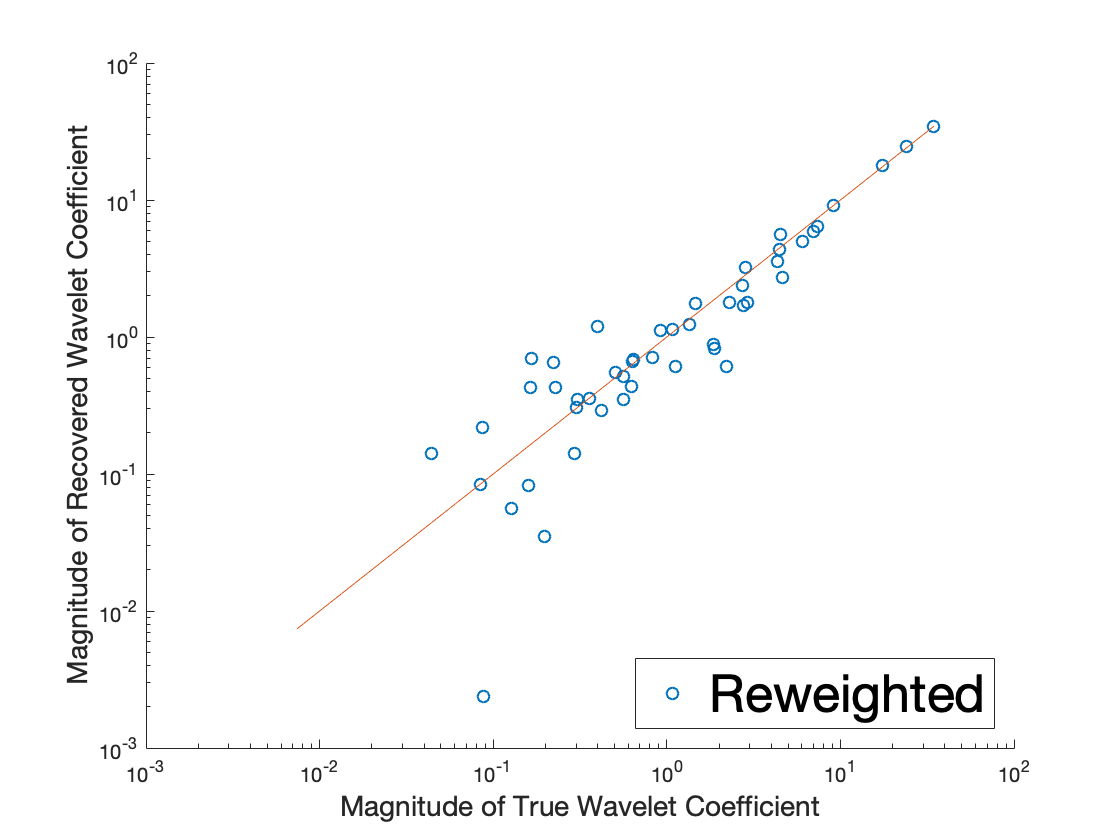}
        \caption{}
        \label{fig:ct_mag_w_1}
    \end{subfigure}
    ~
    \begin{subfigure}[b]{0.45\textwidth}
        \includegraphics[width=\textwidth]{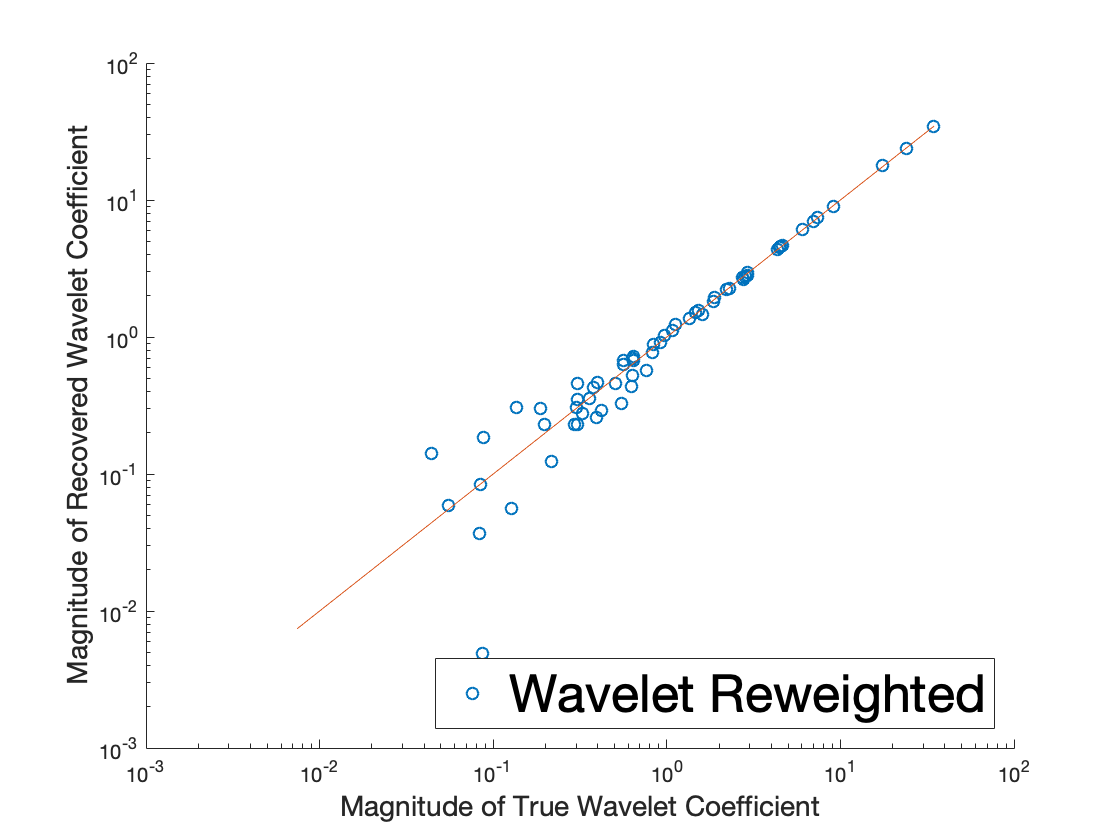}
        \caption{}
        \label{fig:ct_mag_w_threehalf}
    \end{subfigure}
    \caption{A series of plots of the magnitude of the recovered coefficients on the vertical axis and the true
    coefficient on the horizontal axis for various choice of weights.} 
    \label{fig:closed_tree_coef_recov_magnitude}
\end{figure*}

\begin{figure*} 
    \centering
    \begin{subfigure}[b]{0.45\textwidth}
        \includegraphics[width=\textwidth]{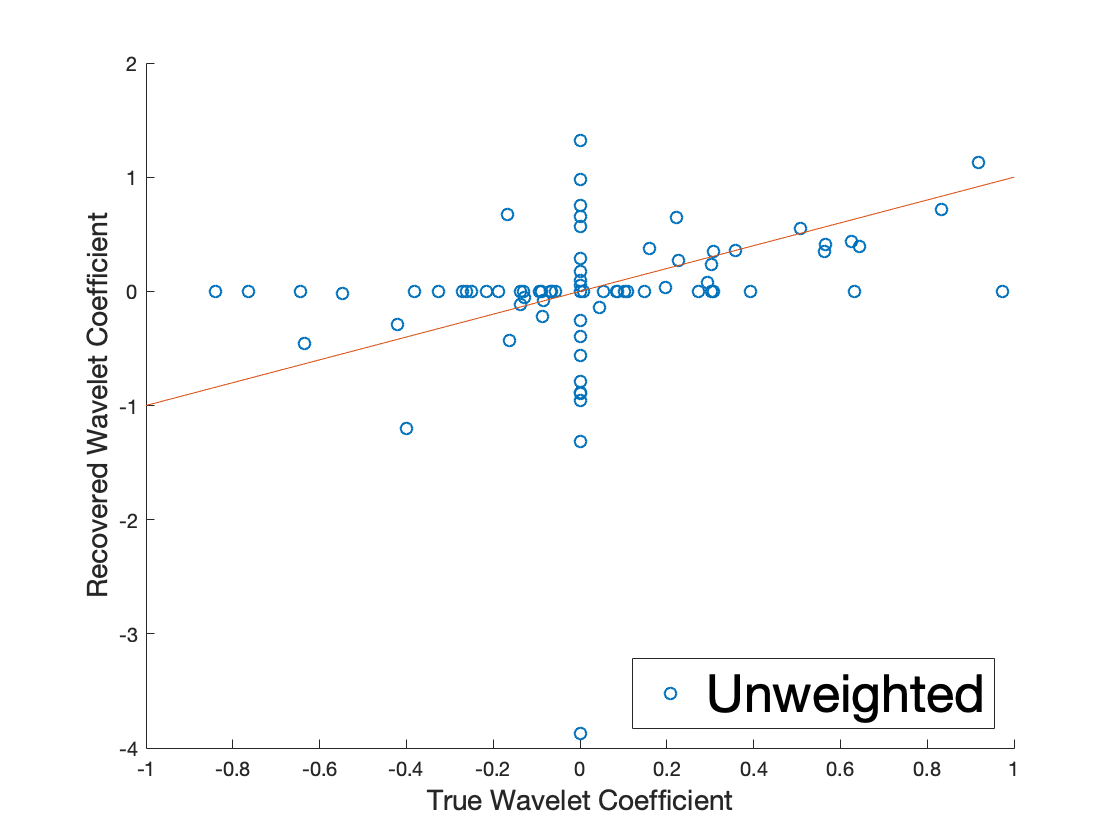}
        \caption{}
        \label{fig:ct_unweighted}
    \end{subfigure}
    ~ 
    \begin{subfigure}[b]{0.45\textwidth}
        \includegraphics[width=\textwidth]{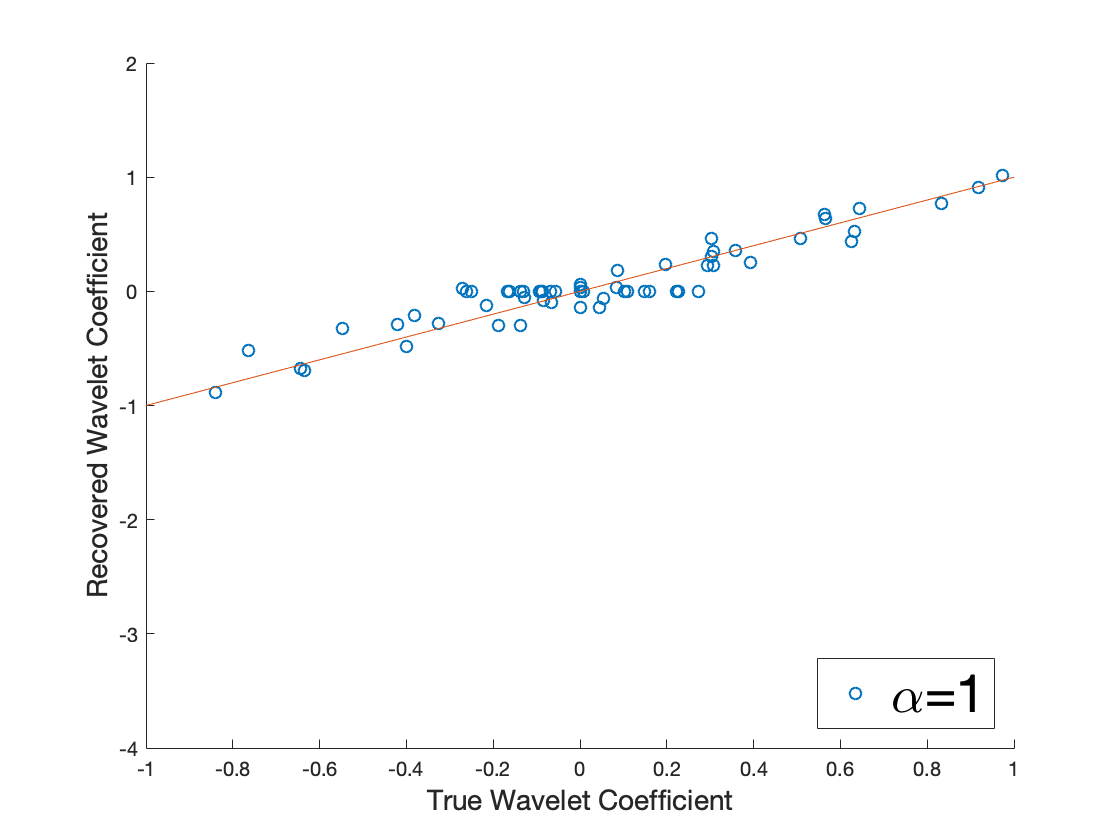}
        \caption{}
        \label{fig:ct_w_half}
    \end{subfigure}
    
    \begin{subfigure}[b]{0.45\textwidth}
        \includegraphics[width=\textwidth]{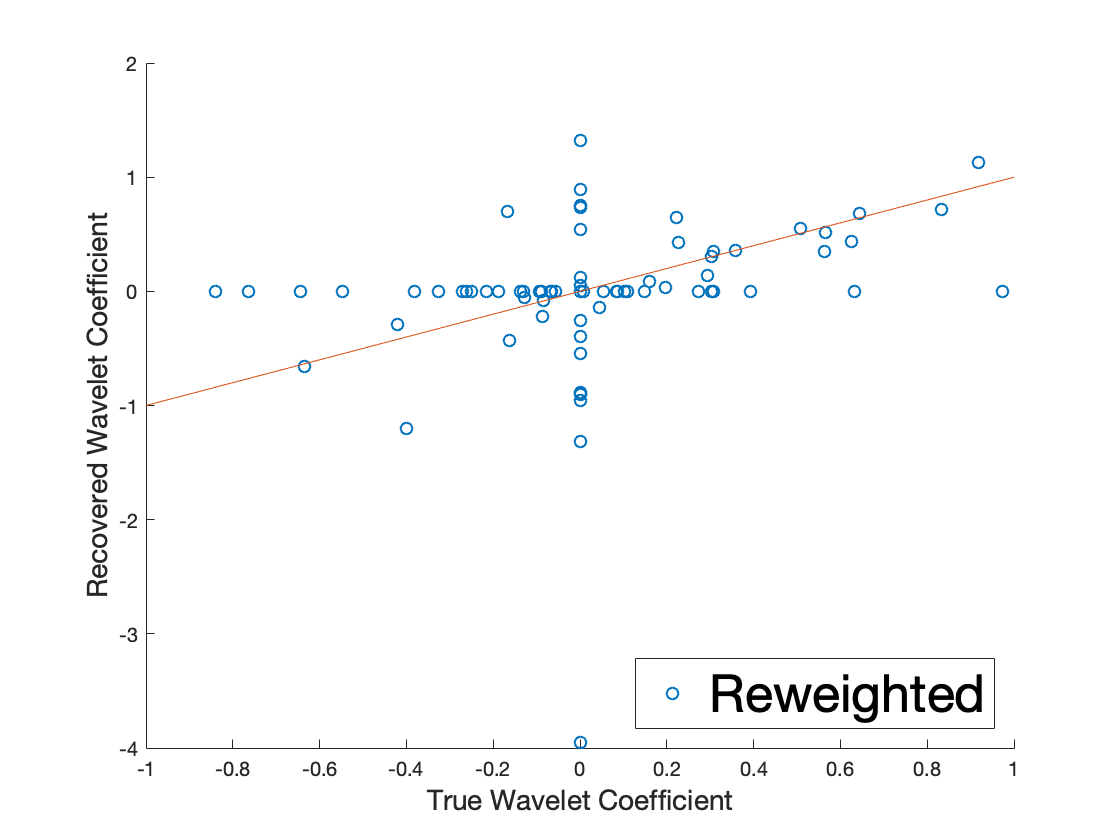}
        \caption{}
        \label{fig:ct_w_1}
    \end{subfigure}
    ~
    \begin{subfigure}[b]{0.45\textwidth}
        \includegraphics[width=\textwidth]{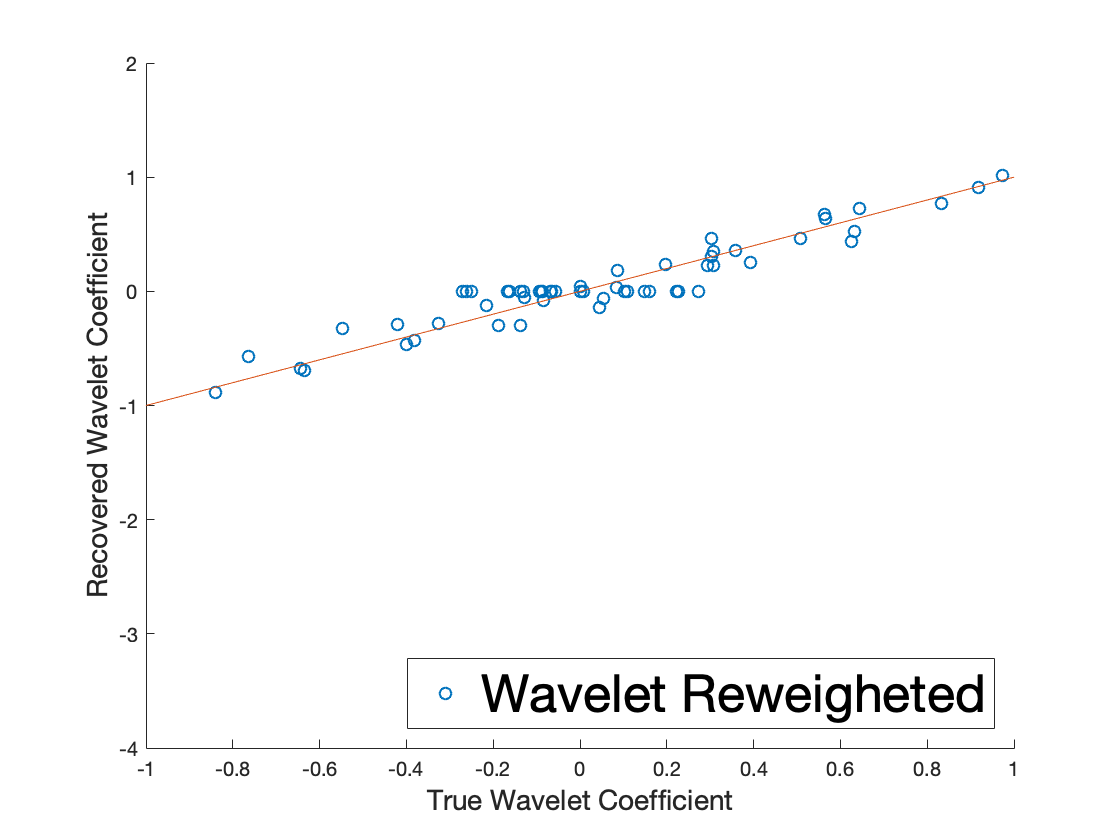}
        \caption{}
        \label{fig:ct_w_threehalf}
    \end{subfigure}
    \caption{A series of plots of the recovered coefficients which correspond to true coefficients on the interval $[-1,1]$ with the
    value of the recovered coefficient on the vertical axis and the true value of the coefficient on the horizontal axis 
    for various choice of weights.} 
    \label{fig:closed_tree_coef_recov}
\end{figure*}

Figure \ref{fig:closed_tree_coef_recov_magnitude} and Figure \ref{fig:closed_tree_coef_recov} 
compare the recovery of a randomly generated closed tree with $90$ nodes which is a 
subtree of wavelet tree with $2^9-1=511$ total nodes using
weighted and unweighted $\ell_1$-minimization. In each of the Figures, the recovered coefficients are
associated with the vertical axis and the true coefficients are associated with
the horizontal axis. If exact recovery is achieved then the points should all lie on the red
line. Using a random sample of $m = 179$ evaluations, we see that unweighted, 
weighted with $\alpha < 1$, and reweighted $\ell_1$-minimization identifies the significant 
coefficents. This can be seen in Figure \ref{fig:closed_tree_coef_recov_magnitude}, where the magnitudes
of the recovered coefficients are plotted. Notice that the weighted approach is better able to capture the 
small coefficients. This is highlighted by Figure \ref{fig:closed_tree_coef_recov} where
we plot the recovered coefficients against the true coefficietns in the interval $[-1,1]$. 

Real-world signals and images do not possess wavelet coefficients which are exactly sparse
and the large coefficients are unlikely exactly closed trees. Rather,
they are often compressible in a wavelet basis. In this section we show that
signals and images can be recovered from a relatively small number of measurements
using weighted $\ell_1$-minimization for the specific choice of weights
\eqref{eq:choice_of_weight_alpha_pos}. Our numerical experiments show that 
for $\alpha = 1$, weighted $\ell_1$-minimization far outperforms both unweighted 
$\ell_1$-minimization and the usual reweighted $\ell_1$-minimization.

\begin{figure*} 
    \centering
    \begin{subfigure}[b]{0.3\textwidth}
        \includegraphics[width=\textwidth]{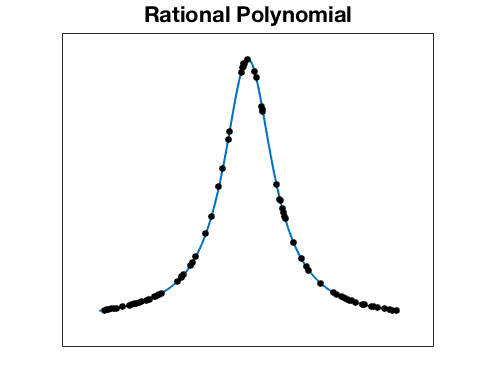}
        \caption{}
        \label{fig:rational_poly}
    \end{subfigure}
    ~ 
    \begin{subfigure}[b]{0.3\textwidth}
        \includegraphics[width=\textwidth]{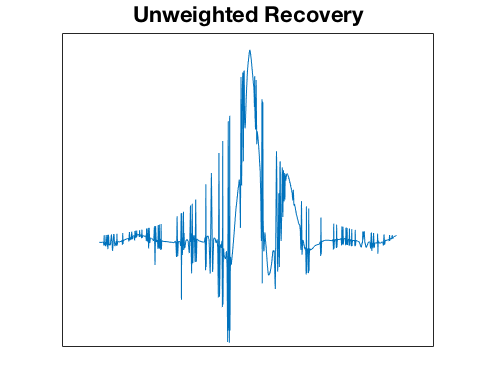}
        \caption{}
        \label{fig:rational_poly_unweighted}
    \end{subfigure}
    ~ 
    \begin{subfigure}[b]{0.3\textwidth}
        \includegraphics[width=\textwidth]{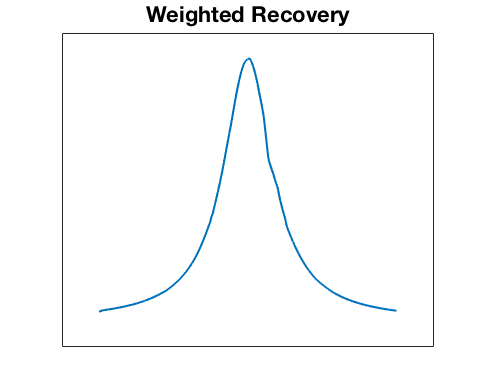}
        \caption{}
        \label{fig:rational_poly_weighted}
    \end{subfigure}
    \caption{Reconstruction of a rational polynbomial using unweighted and weighted $\ell_1$-minimization.  Here we plot:
    in Figure 
    (\ref{fig:rational_poly}) a plot of the rational polynomial $f(x) = 1/(1 + 25x^2)$. 
                  The black dots indicate 80 randomly subsampled values; 
                 in Figure  (\ref{fig:rational_poly_unweighted}) Reconstruction using unweighted 
                  $\ell_1$-minimization and 80 subsampled values; and in Figure (\ref{fig:rational_poly_weighted}) Reconstruction using weighted  $\ell_1$-minimization and 80 subsampled values.
                  } 
    \label{fig:rational_poly_recovery}
\end{figure*}

Figure \ref{fig:rational_poly_recovery} compares the recovery of the {function} $1/(1+25x^2)$ 
from $80$ {uniformly} subsampled points chosen in the interval $[-1,1]$ for different values of $\alpha$ 
from \eqref{eq:choice_of_weight_alpha_pos} as well as unweighted and reweighted $\ell_1$-minimization. 
The chosen wavelets are the one-dimensional \textit{coiflets} constructed in \cite{coiflet_construction}. 
The black dots in Figure \ref{fig:rational_poly} are the sampling points used in the reconstruction. 
Notice that the function recovered by our weighted approach is better than the one obtained
using the unweighted approach. To quantify this, we calculated the Root-mean-square-error (RMSE) in each case. 
The unweighted case produced an RMSE of $0.3100$ where as the the weighted case produced an 
RMSE of $0.0072$.

\begin{figure*} 
    \centering
    \begin{subfigure}[b]{0.3\textwidth}
        \includegraphics[width=\textwidth]{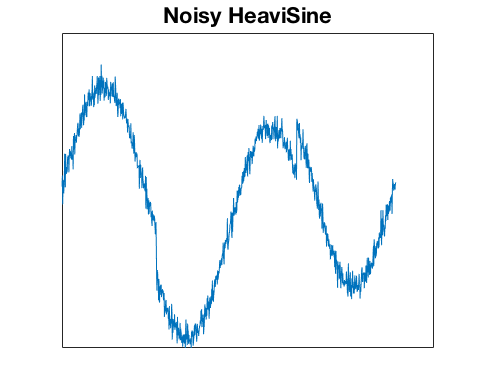}
        \caption{}
        \label{fig:noise_heavisine}
    \end{subfigure}
        ~ 
    \begin{subfigure}[b]{0.3\textwidth}
        \includegraphics[width=\textwidth]{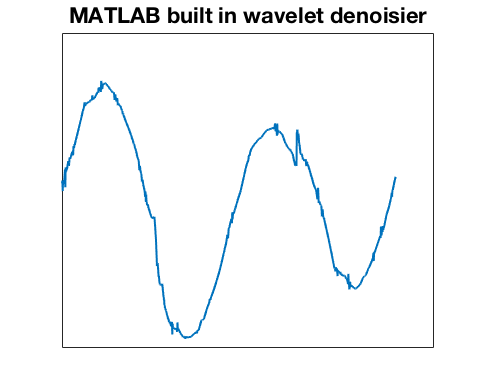}
        \caption{}
        \label{fig:matlab_denoise}
    \end{subfigure}
    ~ 
    \begin{subfigure}[b]{0.3\textwidth}
        \includegraphics[width=\textwidth]{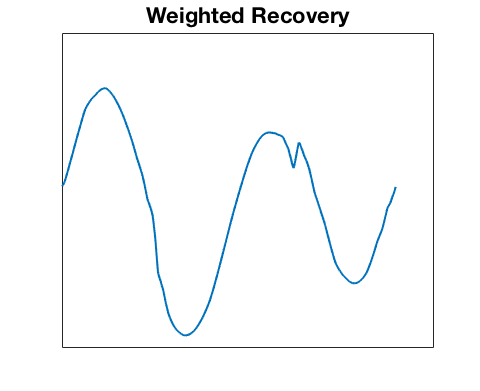}
        \caption{}
        \label{fig:weighted_denoise}
    \end{subfigure}
    \caption{Denoising a perturbed HeaviSine function. Here we plot:  in Figure
    (\ref{fig:noise_heavisine}) the HeaviSine function perturbed by noise; in Figure
                  (\ref{fig:matlab_denoise}) denoised using db3 based wavelet thresholding with the
                  built in matlab function \texttt{wden}; and in Figure 
                  (\ref{fig:weighted_denoise}) Denoised using db3 based weighted 
                  $\ell_1$-minimization.
                  } 
    \label{fig:denoise_1d}
\end{figure*}

We compare two denoising schemes in Figure \ref{fig:denoise_1d}. 
A Gaussian noise was added to the piecewise smooth function as shown 
in Figure \ref{fig:noise_heavisine} so that the PSNR between 
the original Heavisine function and the noisy one is $26.0184$. 
Figure \ref{fig:matlab_denoise} shows the reconstruction using the built-in 
\texttt{MATLAB} function \texttt{wden} which automatically denoises using
the adaptive wavelet shrinkage of the work \cite{wavelet_shrinkage}.
This produces a reconstruction with
PSNR $ = 29.2454$. Figure \ref{fig:weighted_denoise} shows the reconstruction using our
proposed weighted $\ell_1$-minimization scheme and the PSNR is $27.6637$.
While the built-in \texttt{MATLAB} function \texttt{wden} yields a reconstruction with better
PSNR, notice that our reconstruction is more faithful to the 
features of the original signal and does not exhibit the extraneous fluctuations seen 
in Figure \ref{fig:matlab_denoise}. 

\subsection{Recovery of Images}
In this section we consider the problem of reconstructing images from a small percentage of 
its pixels. In the RGB color model, the pixels of images are associated with 3-tuple describing 
a color. Images may be recovered by solving the multiple measurement vectors 
{(MMV)} version of weighted $\ell_1$-minimization, i.e., we solve 
	\begin{equation} \label{eq:mmv_weighted_l1_min}
		\min_{\bm{C} \in \CC^{N \times k}} \lambda \| \bm{C} \|_{\bm{\omega},1,2} 
		+ \| \bm{A} \bm{C} - \tilde{\bm{F}} \|_{F}^2,
	\end{equation}
where $\| \bm{C} \|_{\bm{\omega},1,2}$ is a mixed norm defined as the weighted sum of the $\ell_2$-norms
of the rows of the $N \times k$ matrix $\bm{C}$, $\tilde{\bm{F}}$ is a $m \times 3$ matrix whose columns are
the normalized observations of $f$ along each color band and where $\| \cdot \|_F$ is the Frobenius norm. 

\begin{figure*} 
    \centering
    \begin{subfigure}[b]{0.23\textwidth}
        \includegraphics[width=\textwidth]{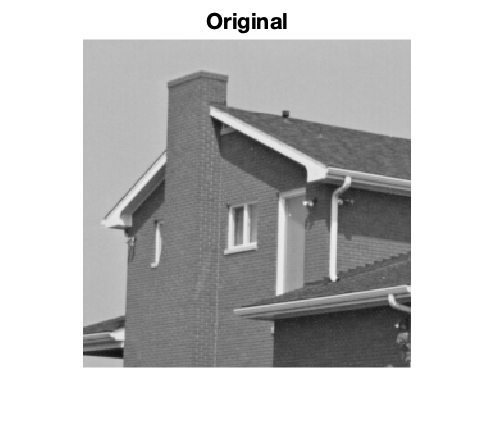}
        \caption{}
        \label{fig:house_orig}
    \end{subfigure}
    ~ 
    \begin{subfigure}[b]{0.23\textwidth}
        \includegraphics[width=\textwidth]{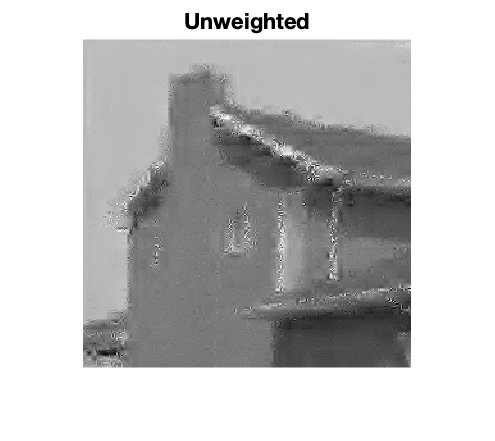}
        \caption{}
        \label{fig:house_unweighted}
    \end{subfigure}
    ~ 
    \begin{subfigure}[b]{0.23\textwidth}
        \includegraphics[width=\textwidth]{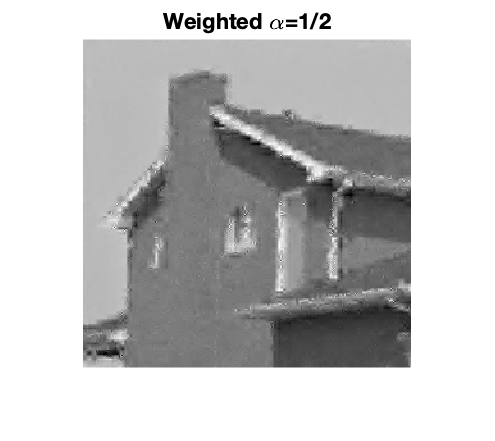}
        \caption{}
        \label{fig:house_alpha12}
    \end{subfigure}
    ~
    \begin{subfigure}[b]{0.23\textwidth}
        \includegraphics[width=\textwidth]{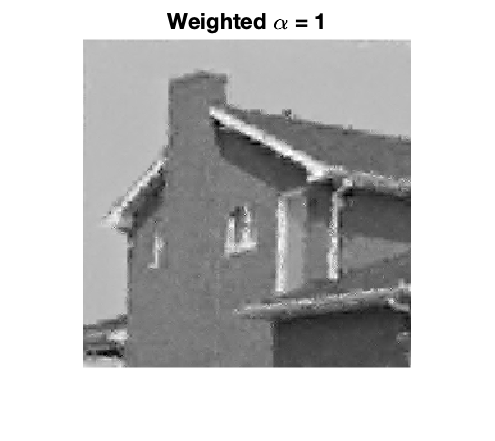}
        \caption{}
        \label{fig:house_alpha1}
    \end{subfigure}

    \begin{subfigure}[b]{0.23\textwidth}
        \includegraphics[width=\textwidth]{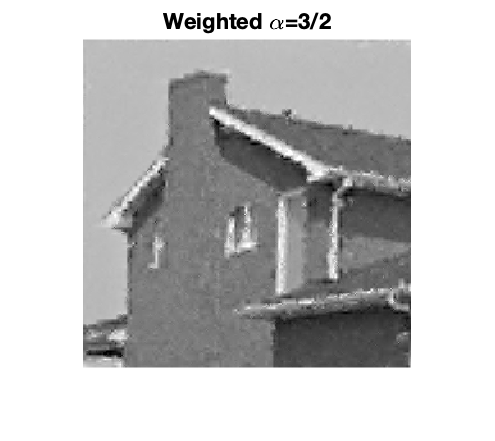}
        \caption{}
        \label{fig:house_32}
    \end{subfigure}
    ~
    \begin{subfigure}[b]{0.23\textwidth}
        \includegraphics[width=\textwidth]{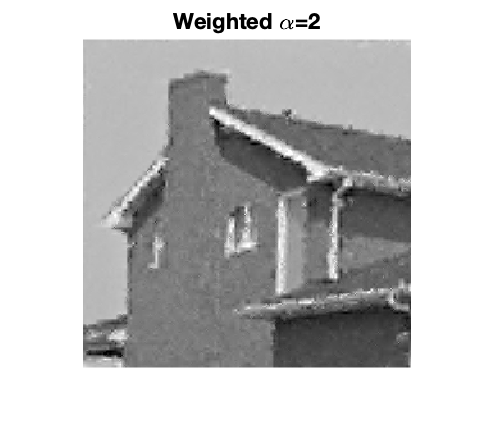}
        \caption{}
        \label{fig:house_2}
    \end{subfigure}
    ~
    \begin{subfigure}[b]{0.23\textwidth}
        \includegraphics[width=\textwidth]{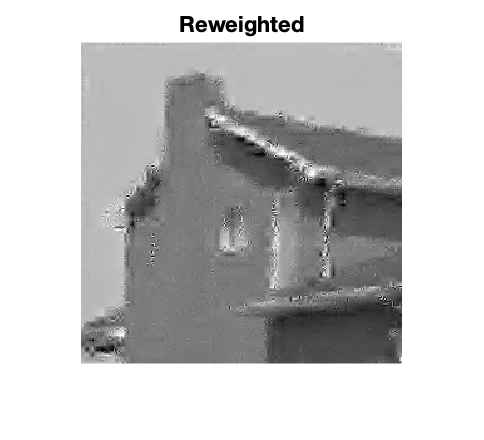}
        \caption{}
        \label{fig:house_rw}
    \end{subfigure}    
    ~
    \begin{subfigure}[b]{0.23\textwidth}
        \includegraphics[width=\textwidth]{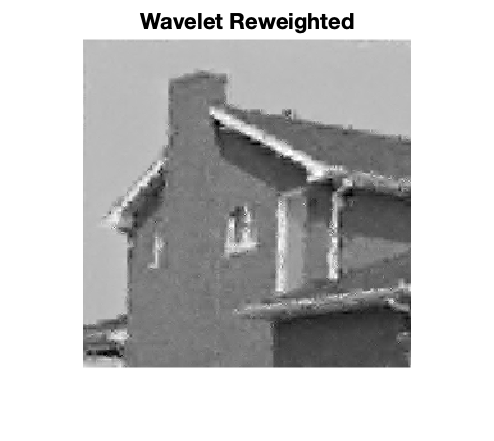}
        \caption{}
        \label{fig:house_nrw}
    \end{subfigure}  
    ~
    \caption{A comparison of the recovered image of a house from $15\%$ randomly chosen pixels 
    using several choices of weights and 
    iterated weight choices. The measurements where taken with respect to the Daubechies 2 (db2) wavelet basis.
    } 
    \label{fig:house_recovery}
\end{figure*}

Figure \ref{fig:house_recovery} shows the recovery of a greyscale house image using several choices of $\alpha$.
The original image has $256 \times 256$ pixels and can be represented in the Haar wavelet basis
with $256^2$ coefficeints. The measurements, $F$, are randomly chosen pixels of the image so that 
$m = 9830$, that is, the measurements are $15\%$ of the $256^2$ pixels, randomly chosen.  
Notice that the cases when $\alpha \ge 1$ vastly out perform IRW $\ell_1$-minimization and unweighted $\ell_1$-minimization.
However, the differences between $\alpha = 3/2$, $\alpha = 2$, and $\alpha = 1$ are minimial. Therefore,
choosing the weights as \eqref{eq:choice_of_weight_alpha} is a reasonable choice in a general situation.

\begin{figure*} 
    \centering
    \begin{subfigure}[b]{0.23\textwidth}
        \includegraphics[width=\textwidth]{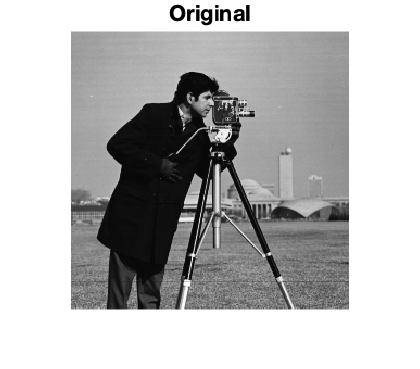}
        \caption{}
        \label{fig:cameraman_orig}
    \end{subfigure}
    ~ 
    \begin{subfigure}[b]{0.23\textwidth}
        \includegraphics[width=\textwidth]{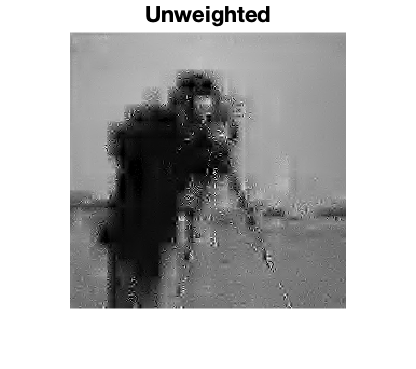}
        \caption{}
        \label{fig:cameraman_unweighted}
    \end{subfigure}
    ~ 
    \begin{subfigure}[b]{0.23\textwidth}
        \includegraphics[width=\textwidth]{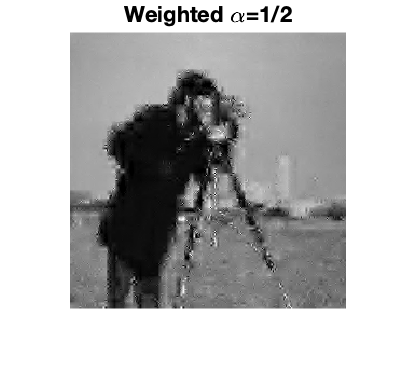}
        \caption{}
        \label{fig:cameraman_alpha12}
    \end{subfigure}
    ~
    \begin{subfigure}[b]{0.23\textwidth}
        \includegraphics[width=\textwidth]{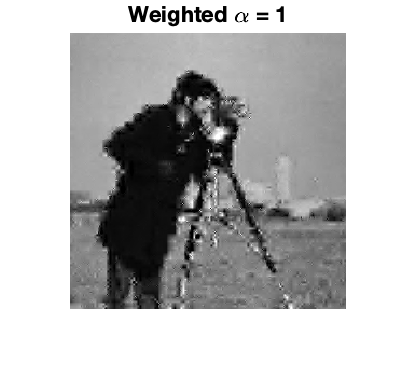}
        \caption{}
        \label{fig:cameraman_alpha1}
    \end{subfigure}

    \begin{subfigure}[b]{0.23\textwidth}
        \includegraphics[width=\textwidth]{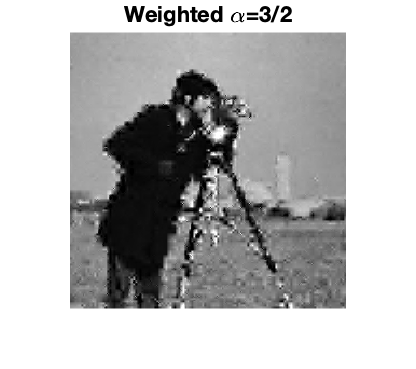}
        \caption{}
        \label{fig:cameraman_32}
    \end{subfigure}
    ~
    \begin{subfigure}[b]{0.23\textwidth}
        \includegraphics[width=\textwidth]{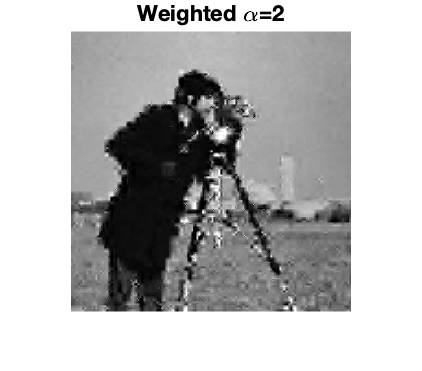}
        \caption{}
        \label{fig:cameraman_2}
    \end{subfigure}
    ~
    \begin{subfigure}[b]{0.23\textwidth}
        \includegraphics[width=\textwidth]{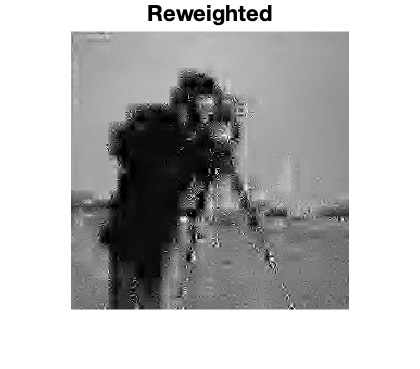}
        \caption{}
        \label{fig:cameraman_rw}
    \end{subfigure}    
    ~
    \begin{subfigure}[b]{0.23\textwidth}
        \includegraphics[width=\textwidth]{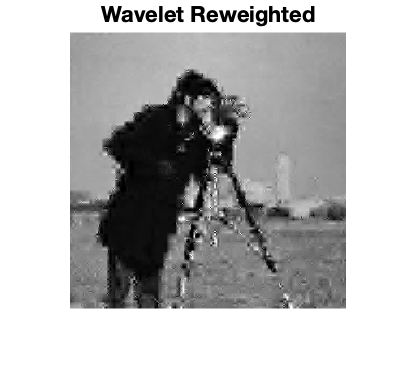}
        \caption{}
        \label{fig:cameraman_nrw}
    \end{subfigure}  
    ~
    \caption{A comparison of the recovered image of a cameraman for a subsample of $10\%$ randomly chosen pixels
    using several choices of weights and 
    iterated weight choices. The measurements where taken with respect to the Daubechies 2 (db2) wavelet basis.
    } 
    \label{fig:cameraman_recovery}
\end{figure*}

    \begin{figure*} 
        \centering
        \begin{subfigure}[b]{0.3\textwidth}
            \includegraphics[width=\textwidth]{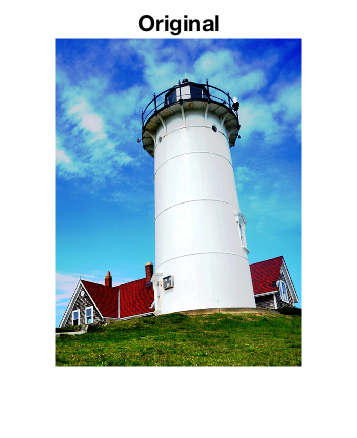}
            \caption{}
            \label{fig:lighthouse}
        \end{subfigure}
        ~ 
        \begin{subfigure}[b]{0.3\textwidth}
            \includegraphics[width=\textwidth]{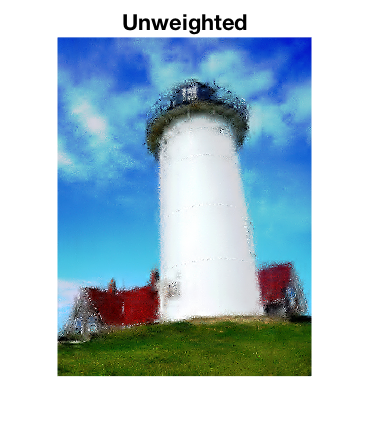}
            \caption{}
            \label{fig:lighthouse_unweighted}
        \end{subfigure}
        ~ 
        \begin{subfigure}[b]{0.3\textwidth}
            \includegraphics[width=\textwidth]{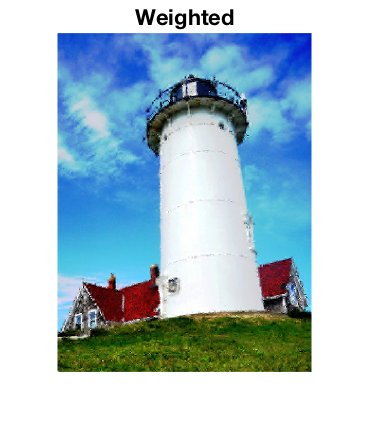}
            \caption{}
            \label{fig:lighthouse_weighted}
        \end{subfigure}
        \caption{Subsampling reconstruction of an image with unweighted and weighted $\ell_1$-minimization. Here we plot: in Figure (\ref{fig:lighthouse}) the original $640 \times 480$ pixel image of a lighthouse; in Figure (\ref{fig:lighthouse_unweighted}) Reconstruction using db3 based unweighted 
                      $\ell_1$-minimization and $15$ \% randomly subsampled pixels; and in Figure (\ref{fig:lighthouse_weighted}) Reconstruction using db3 based weighted 
                      $\ell_1$-minimization and $15$ \% randomly subsampled pixels. 
                      } 
        \label{fig:lighthouse_recovery}
    \end{figure*}

We can also recover color images by solving the minimization problem 
\eqref{eq:mmv_weighted_l1_min}. Figure \ref{fig:lighthouse_recovery} shows that 
the weighted approach performs better than unweighted for color images. The PSNR
of the reconstruction using unweighted $\ell_1$-minimization is $21.3119$, 
see Figure \ref{fig:lighthouse_unweighted}. On the other hand, 
the PSNR using weighted $\ell_1$-minimization is $24.5694$, see Figure \ref{fig:lighthouse_weighted}.
Notice that the unweighted recovery features blurring of edges and does not recover the texture 
of either the grass or the red roof tile. The weighted recovery exhibits a better recovery of sharp edges and the
texture of the grass with yellow flowers. Weighted $\ell_1$-minimization can also be deployed to 
recover other kinds of images besides the ``natural landscape" type images typified by the lighthouse. 
Below we consider recovering cartoons, textures, and scientific data.

\begin{figure*} 
    \centering
    \begin{subfigure}[b]{0.3\textwidth}
        \includegraphics[width=\textwidth]{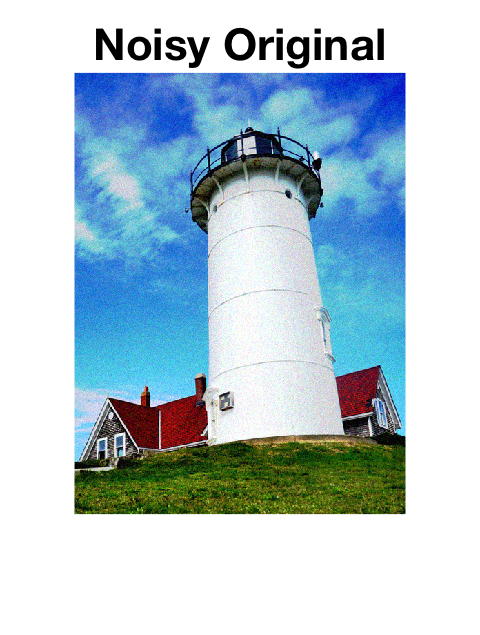}
        \caption{}
        \label{fig:noisy_lighthouse}
    \end{subfigure}
        ~ 
    \begin{subfigure}[b]{0.3\textwidth}
        \includegraphics[width=\textwidth]{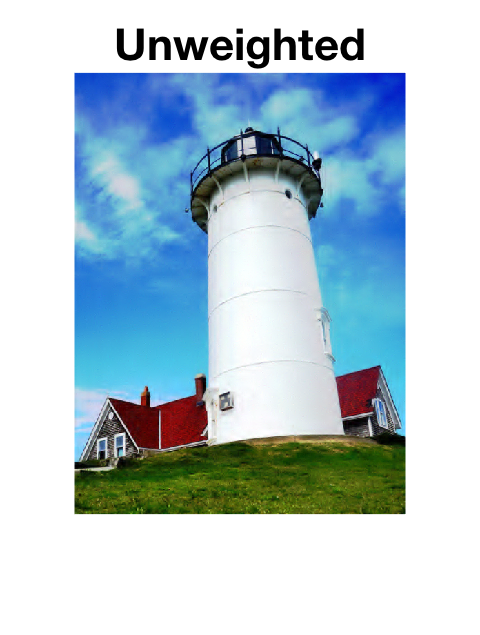}
        \caption{}
        \label{fig:unweighted_denoised_lighthouse}
    \end{subfigure}
    ~ 
    \begin{subfigure}[b]{0.3\textwidth}
        \includegraphics[width=\textwidth]{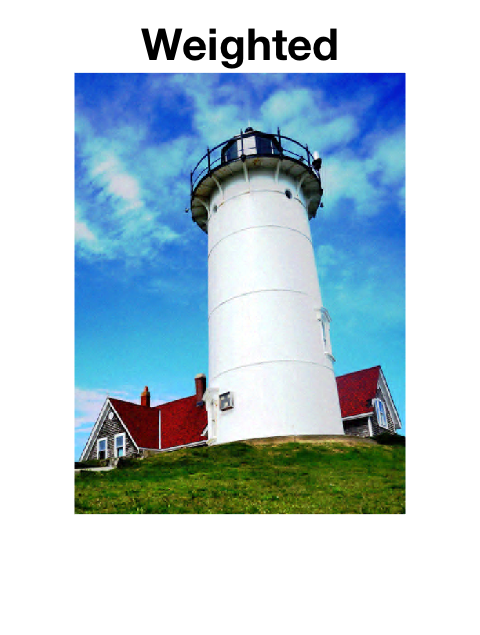}
        \caption{}
        \label{fig:weighted_denoised_lighthouse}
    \end{subfigure}
    \caption{Denoising an image with unweighted and weighted $\ell_1$-minimization.  Here we plot: in Figure
    (\ref{fig:noisy_lighthouse}) A image of a lighthouse with noise; in Figure 
            (\ref{fig:unweighted_denoised_lighthouse}) Denoised using db3 based unweighted
                  $\ell_1$-minimization; and in Figure
        (\ref{fig:weighted_denoised_lighthouse}) Denoised using db3 based weighted 
                  $\ell_1$-minimization. 
                  } 
    \label{fig:denoise_2d}
\end{figure*}

We also present an example of image denoising. Figure \ref{fig:noisy_lighthouse} is a noisy
image generated by adding a Gaussian noise so that the PSNR of the noisy version is 26.0184. The reconstruction obtained using unweighted $\ell_1$-minimization
has PSNR $=30.6720$, see Figure \ref{fig:unweighted_denoised_lighthouse},
and the weighted $\ell_1$-minimization reconstruction has a PSNR $=31.1165$,
see Figure \ref{fig:weighted_denoised_lighthouse}.

\subsection{Recovering Hyperspectral Images}
The pixels of the color images we recovered in the previous section 
can be viewed as $3$-tuples of numbers which represent the color at each pixel. 
The image itself can then be viewed as an object in $\mathbb{R}^{M \times N \times 3}$
where $M$ is the number of pixel along the width and $N$ is the number
of pixels along the length. 
A hyperspectral image is an object in 
$\mathbb{R}^{M \times N \times k}$ for some $k>1$ where
$M$ and $N$ are the spatial dimensions and $k$ is the number of spectral bands. 
One can use the information stored in a hyperspectral image in a variety of contexts. 
Frequently, hyperspectral images
are used for the remote detection or classification \cite{hyper_image}. In particular,
it has been used in medicine \cite{hyper_medicine} for detection and classification
of disease, and geology \cite{hyper_geo} for detection and classification of minerals or oil.

In our numerical experiment, we consider recovering a hyperspectral image from a set of 
subsampled spectral profiles at $m$ randomly chosen locations. In other words, we sample
$m$ vectors $\mu_{i,j} \in \mathbb{R}^k$ from the hyperspectral image 
and wish to recover the full tensor. We do this by solving \eqref{eq:mmv_weighted_l1_min}.
For our experiment we have used a hyperspectral image associated with 
a natural landscape of fields. The spectrum at each pixel corresponds to 
the presense of certain wavelengths of light. For a sample of the spectral profiles
at $25\%$ of the pixels we recover the tensor using weighted and unweighted 
$\ell_1$-minimization.

\begin{figure*}
    \centering
        \includegraphics[width=\textwidth]{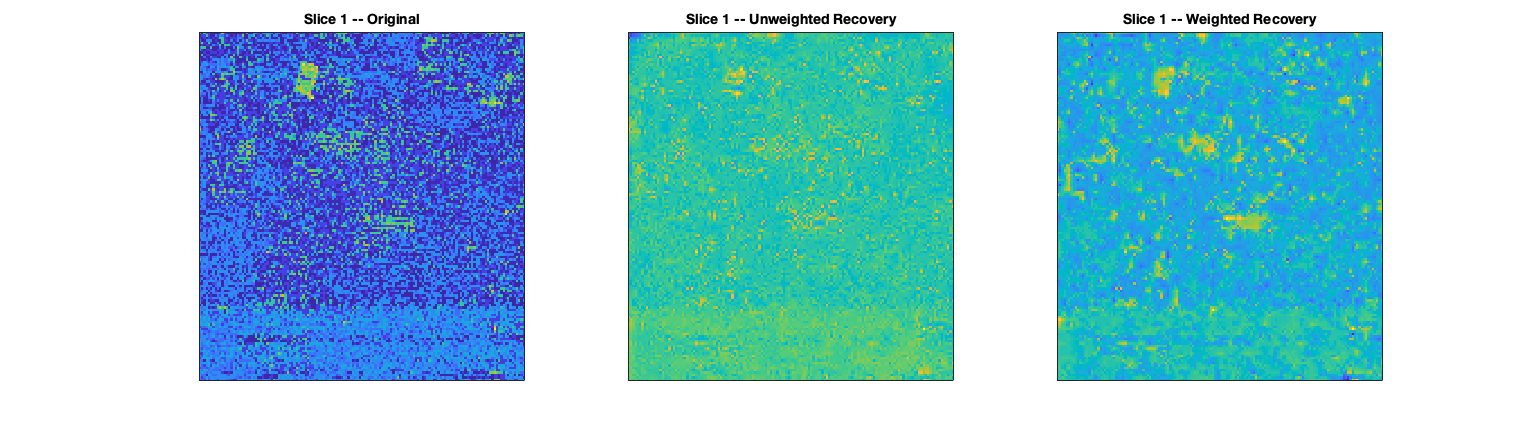}
        \caption{A comparison of  the recovered slices at the first spectral index.}
        \label{fig:hyper_slice_1}
\end{figure*}

\begin{figure*}
    \centering
        \includegraphics[width=\textwidth]{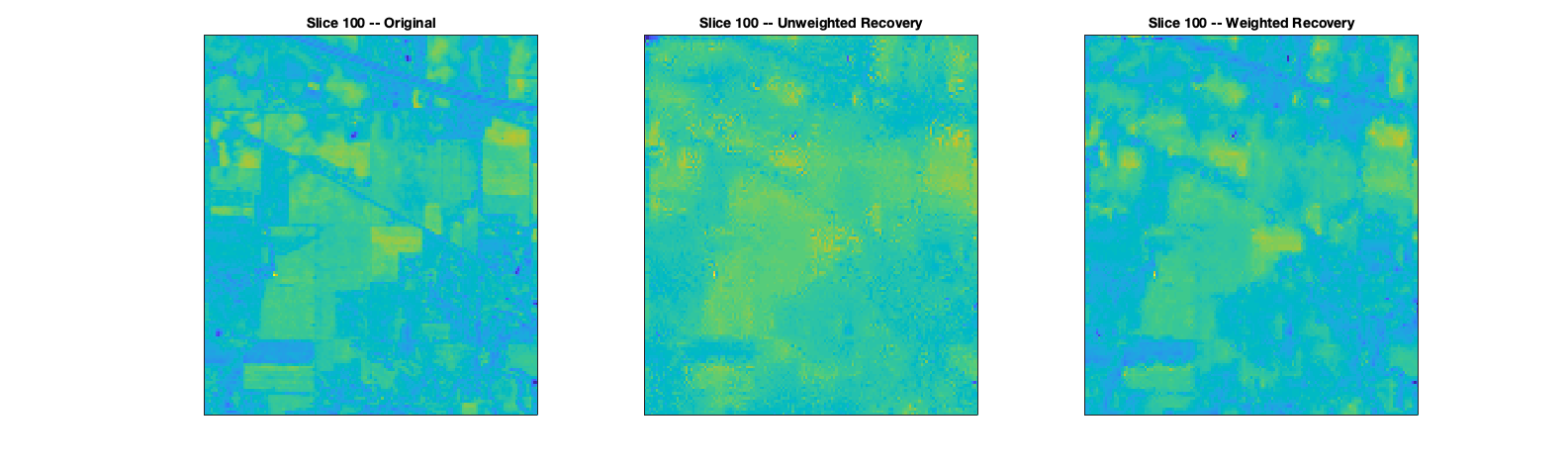}
        \caption{A comparison of the recovered slices at the $100^{th}$ spectral index.}
        \label{fig:hyper_slice_100}
\end{figure*}

\begin{figure}
    \centering
    \includegraphics[width=0.49\textwidth]{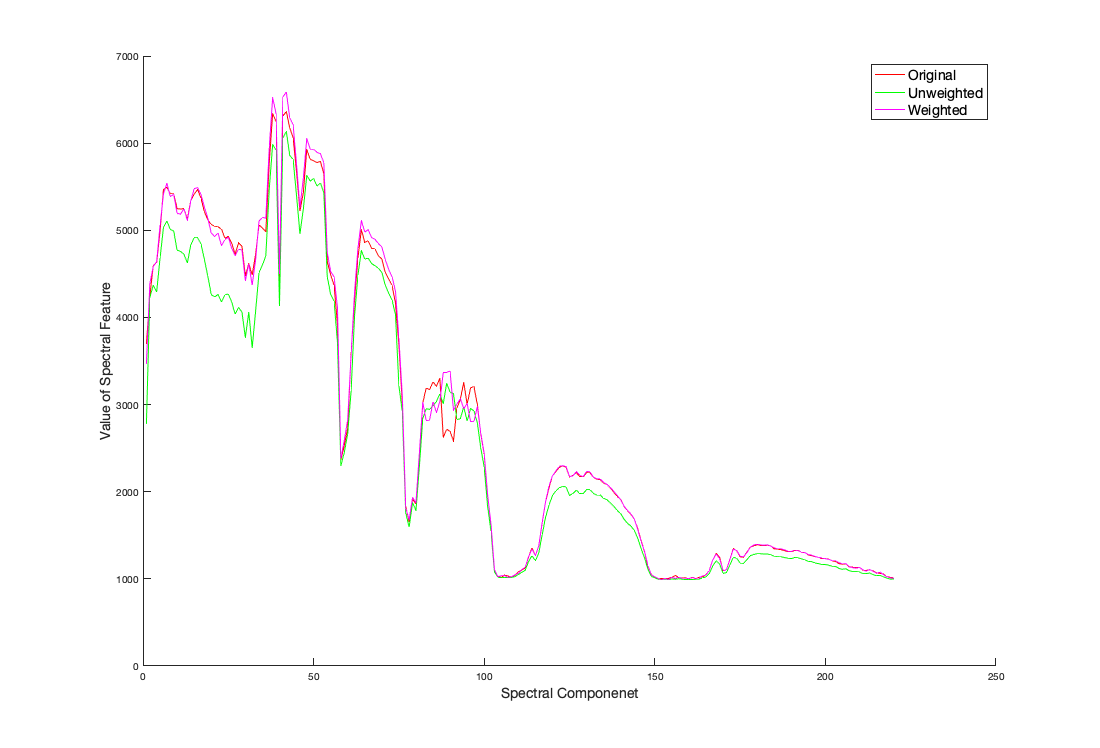}
    \caption{A comparison of the recovered slices at the $100^{th}$ spectral index.}
    \label{fig:hyper_spec_pro}
\end{figure}

In Figure \ref{fig:hyper_slice_1} and Figure \ref{fig:hyper_slice_100} we compare
recovered slices of the tensor at spectral index $1$ and spectral index $100$ 
respectively. Notice that the unweighted approach does not yield as good results
as the weighted approach.
For a particular pixel we can compare the recovery by looking at the spectral profile
associated with that pixel. The spectral profile for the pixel $(50,25)$ and the recovered versions
are plotted in Figure \ref{fig:hyper_spec_pro}. 

\subsection{Haar Framelets} \label{sec:framelets}
Many successful image processing methods incorporate both local 
and global information about a signal to increase performance
\cite{nonlocal_means,morel_denoising2005,Kheradmand2014,LDMM}. 
In this section we consider a specific case of a representation system introduced in 
\cite{a_tale_of_two_bases} where the simultaneous local and global feature
analysis of an OoI is performed by a dictionary called a
\textit{framelet}. A sparse representation in the framelet dictionary 
recovered from a subsample set of measurements using our proposed 
weighted $\ell_1$-minimization problem. The dictionary is constructed 
by taking the convolution of so called ``local" and ``global" 
bases discussed in more detail below. 

Let $\bm{F} =(F_0,F_1,\dots,F_{N-1})\in \R^N$ be the vector representing
the target digital signal. Local information is gathered by grouping
neighboring evaluations around every point
together into an array called a \textit{patch}. For each $k$, 
$0\leq k< N$, the patch  of length $\ell$ at location $k$ is defined
$\bm{p}_{k} = (F_k,F_{k+1}, \dots,F_{k+\ell-1})$ where $k+\ell-1$ is
interpreted as  circular addition, i.e. $(N-1)+1$ is identified with $0$,
$(N-1)+2$ is identified with $1$ and so on. 
The \textit{patch matrix} $P$ is constructed by setting the vector $\bm{p}_{k}$ as 
the $k^{th}$ row of $P$. Notice that $P$ has $N$ rows, one for each
value in $\bm{F}$, and $\ell$ columns corresponding to the patch size. 

The global basis is given as a matrix $G \in \R^{N \times N}$ with its columns forming
an orthonormal basis in $\R^N$, and the local basis is given as a matrix 
$L \in \R^{\ell \times \ell}$ with its columns forming 
an orthonormal basis in $\R^{\ell}$. 
The patch matrix $P$ can be represented in the tensor product 
basis genereated from $G$ and $L$ with the coefficients computed by 
    \begin{equation}\label{eq:C_coef}
        C=G^TPL.        
    \end{equation}
The entries of the matrix $C=(c_{i,j})$ can also be viewed as coefficients of 
$\bm{F}$ in the convolutional framelet formed by the columns of $G$ and $L$. 
\begin{definition}[Discrete, Circular Convolution]
		For two vectors $\bm{v},\bm{w}$ of length $N$ we define the discrete, 
		circular convolution as an operator which returns a length 
		$N$ vector $(\bm{v} \ast \bm{w})$ whose 
		$k^{th}$ component is 
		\begin{equation}
			(\bm{v} \ast \bm{w})[k] = \sum_{p=0}^{N-1} \bm{v}[k-p] \bm{w}[p]
		\end{equation}
	\end{definition}

Let $\bm{G}_i$ be the $i^{th}$ column of the matrix 
$G$ and $\bm{L}_j$ be the $j^{th}$ column of $L$. Denote by $\bar{\bm{L}}_j$ 
the vector $\R^N$ whose first $l$ entries are identical with corresponding 
entries in $\bm{L}_j$, and the rest are equal to $0$. The convolutional framelets
are constructed as the circular convolution of
$\bm{G}_i$ with $\bar{\bm{L}}_j$:
\begin{equation}
		\bm{\varphi}_{i,j} = \frac{1}{\sqrt{\ell}}\, \bm{G}_i \ast \bar {\bm{L}}_j.
	\end{equation}
The vectors $\bm{\varphi}_{i,j}$	form a Parseval frame in $\R^N$ 
(see \cite{christensen2016introduction} for definitions).
The coefficients $c_{i,j}$ from \eqref{eq:C_coef} satisfy
    \begin{equation*}
        c_{i,j}=\langle \bm{F},\bm{\varphi}_{i,j}\rangle,
    \end{equation*}

and the vector $\bm{F}$ can be recovered by the reconstruction formula
	\begin{equation} \label{eq:recov_f}
		\bm{F} =  \sum_{i=1}^N \sum_{j=1}^\ell c_{ij} \bm{\varphi}_{ij}.
	\end{equation}

We choose the Haar basis for both the global and local basis in our numerical example. Consequently, for the the weights $\omega_{i,j}$ we have
	\begin{equation}
		 \omega_{i,j} :=\|\bm{\varphi}_{i,j}\|_{i,j}= 2^{\gamma_i \lambda_j /2}
	\end{equation}
where $\gamma_i$ is the depth of the node associated with the $i^{th}$ wavelet function whose discretization 
is the $i^{th}$ row of $G$ and where $\lambda_j$ is defined similarly for the Haar basis associated with $L$.

\begin{figure*} 
    \centering
    \begin{subfigure}{0.45\linewidth}
    \centering
        \includegraphics[width=\linewidth]{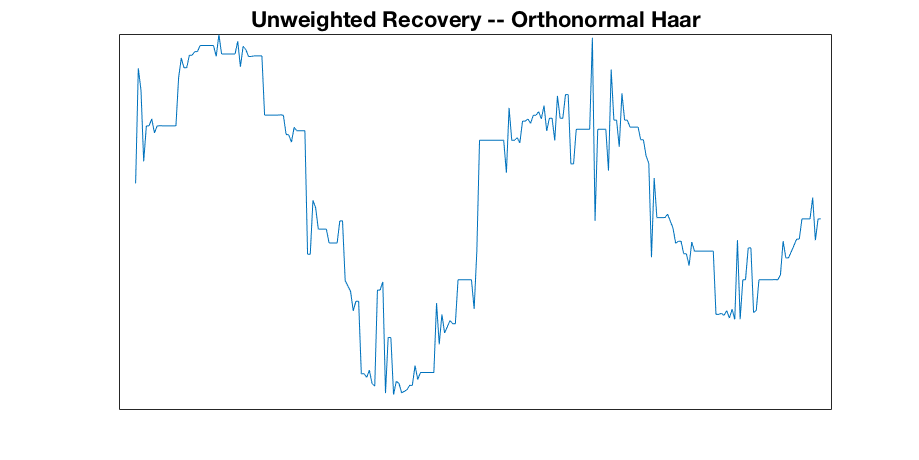}
        \caption{}
        \label{fig:unweighted_ortho}
    \end{subfigure}
    \hfill 
    \begin{subfigure}{0.45\linewidth}
    \centering
        \includegraphics[width=\linewidth]{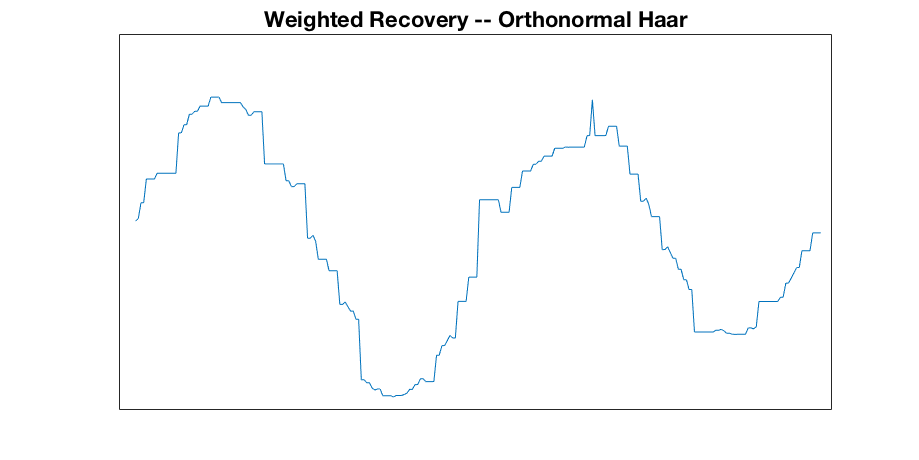}
        \caption{}
        \label{fig:weighted_ortho}
    \end{subfigure}
    
	\bigskip
    \begin{subfigure}{0.45\linewidth}
        \includegraphics[width=\linewidth]{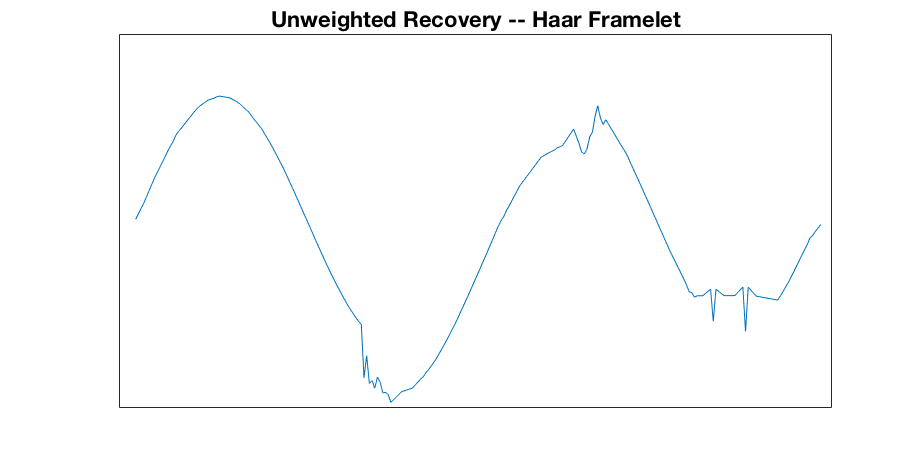}
        \caption{}
        \label{fig:unweighted_framelet}
    \end{subfigure}
    \hfill 
    \begin{subfigure}{0.45\linewidth}
        \includegraphics[width=\linewidth]{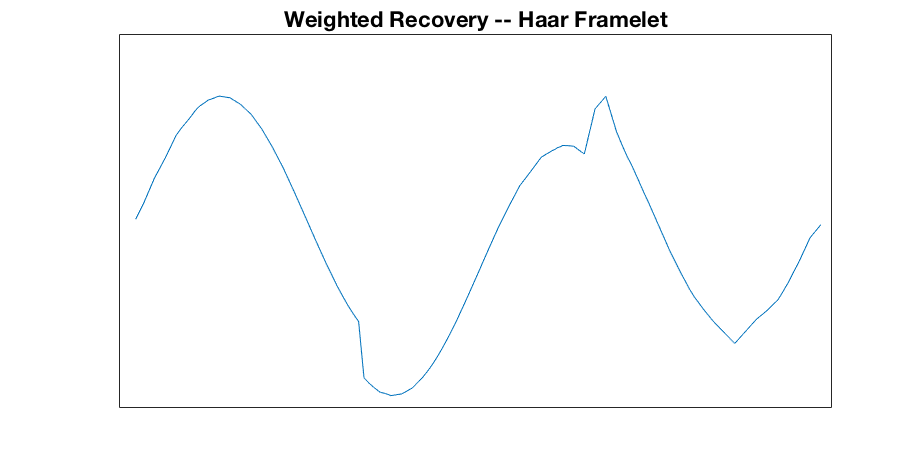}
        \caption{}
        \label{fig:weighted_framelet}
    \end{subfigure}
    \caption{\eqref{fig:unweighted_ortho} The recovery of the HeaviSine function using
                  an orthonormal Haar basis and unweighted $\ell_1$-minimization. 
                  \eqref{fig:weighted_ortho} The recovery of the HeaviSine function using 
                  an orthonormal Haar basis and weighted $\ell_1$-minimization.
                  \eqref{fig:unweighted_framelet} The recovery of the HeaviSine function 
                  using Haar Framelets and unweighted $\ell_1$-minimization.
                  \eqref{fig:weighted_framelet} The recovery of the HeaviSine function using
                  Haar Framelets and weighted $\ell_1$-minimization.
                  } 
    \label{fig:framelets_ortho_1d}
\end{figure*}

In Figure \ref{fig:framelets_ortho_1d} 
each of the reconstructions was created from 80 samples of a piecewise smooth function.
Since the function is piecewise smooth, it is not necessary compressible
in the Haar basis. The reconstructions using weighted and unweighted $\ell_1$-minimization
with the orthonormal Haar basis show ``step"-like artifacts. On the other hand, 
the recovered framelet representation does not
exhibit the step-like affects. Heuristically, the observed improved performance may be 
explained by the property that Haar framelets
use local and global information simultaneously. 

\section{Conclusion}
\label{sec:conclusion}
This effort has shown that weighted $\ell_1$-minimization is effective for
solving the interpolation/inpainting and denoising problems by recovering
wavelet coefficients. Moreover, this effort provides two explicit choices for weights that 
do not require the identification of parameters beyond the choice of a wavelet family for use as
a representation system. Provided numerical examples indicate that the choice of weights
\eqref{eq:choice_of_weight_alpha} far outperforms unweighted $\ell_1$-minimization for recovering 
wavelet coefficients and that there is little difference between the case when $\alpha > 1$ and $\alpha = 1$ for
the weights \eqref{eq:choice_of_weight_alpha_pos}, hence, $\alpha = 1$ is a good choice. 
According to Figure \ref{fig:compare_weights}, the weights used in IRW $\ell_1$-minimization are not scaled 
appropriately. Our choice of weights \eqref{eq:wirwl1} both iteratively updates weights so that large coefficients have 
smaller associated weights and ensures that the updated weights do not become too small. We also show 
that weighted $\ell_1$-minimization can be used for measurement systems that do not happen 
to be an orthonormal system, see Section \ref{sec:framelets}. We have a proof which shows that 
the sampling complexity for our weighted $\ell_1$-minimization is no worse than the sampling complexity 
for unweighted $\ell_1$-minimization assuming that the sparse signal satisfies the closed tree assumption.
In future work, it would be interesting to establish sharp estimates associated with wavelet based measurement systems. 
Such a result would theoretically explain the gap in performance between unweighted and weighted $\ell_1$-minimizations for
recovering wavelet coefficients. In this work we mainly consider images and signals. Another interesting
direction to pursue would be to apply our choice of weights for recovering wavelet coefficients of functions
which are solutions to partial differential equations. 

\section*{Acknowledgements}
This material is based upon work supported in part by: the U.S. Department of Energy, Office of Science, Early Career Research Program under award number ERKJ314; U.S. Department of Energy, Office of Advanced Scientific Computing Research under award numbers ERKJ331 and ERKJ345; the National Science Foundation, Division of Mathematical Sciences, Computational Mathematics program under contract number DMS1620280; Scientific Discovery through Advanced Computing (SciDAC) program through the FASTMath Institute under Contract No. DE-AC02-05CH11231; and by the Laboratory Directed Research and Development program at the Oak Ridge National Laboratory, which is operated by UT-Battelle, LLC., for the U.S. Department of Energy under contract DE-AC05-00OR22725.

\appendix
\subsection{Additional Numerical Examples}
\begin{figure*} 
    \centering
    \begin{subfigure}{0.31\linewidth}
    \centering
        \includegraphics[width=\linewidth]{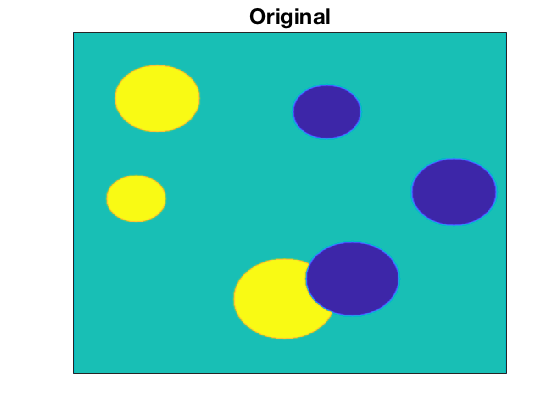}
        \caption{}
        \label{fig:original_circles_cartoon}
    \end{subfigure}
    \begin{subfigure}{0.31\linewidth}
    \centering
        \includegraphics[width=\linewidth]{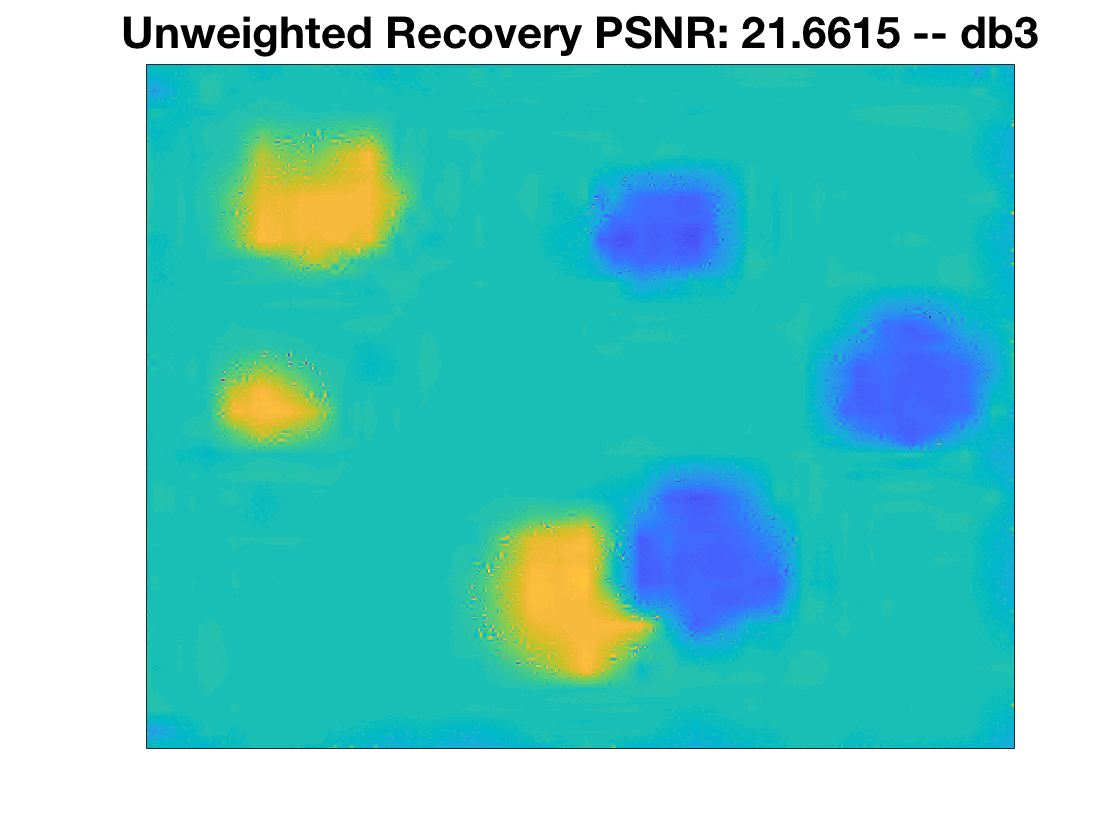}
        \caption{}
        \label{fig:unweighted_circles_cartoon}
    \end{subfigure}
    \begin{subfigure}{0.31\linewidth}
        \includegraphics[width=\linewidth]{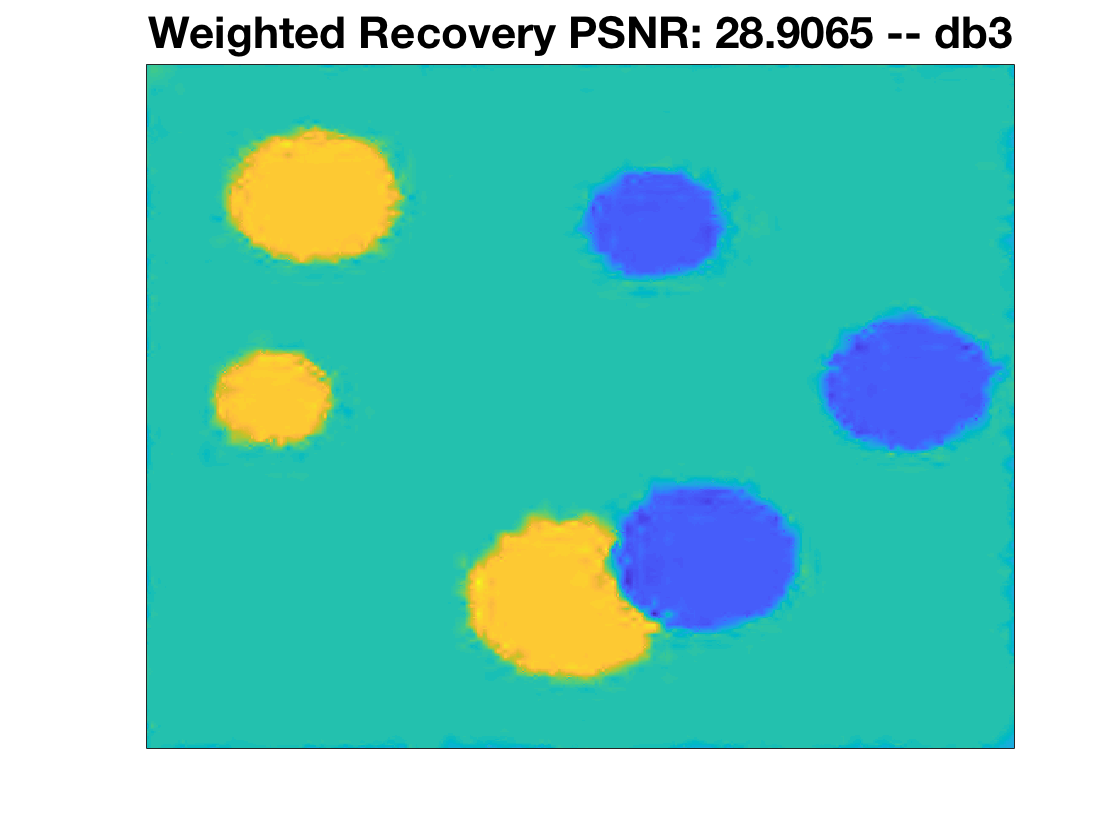}
        \caption{}
        \label{fig:weighted_circles_cartoon}
    \end{subfigure}
    \caption{\eqref{fig:original_circles_cartoon} A $512 \times 512$ pixel cartoon-type image of circles some of which overlap of size. 
                  The image is grayscale 
                  but presented in color for easier viewing. 
                  \eqref{fig:unweighted_circles_cartoon} The recovery of the circles cartoon using unweighted $\ell_1$-minimization and 
                  the db3 basis from 4\% of the pixels.
                  \eqref{fig:weighted_circles_cartoon} The recovery of the circles cartoon using weighted $\ell_1$-minimization and 
                  the db3 basis from 4\% of the pixels.
                  } 
    \label{fig:cartoon}
\end{figure*}

\begin{figure*} 
    \centering
    \begin{subfigure}{0.31\linewidth}
    \centering
        \includegraphics[width=\linewidth]{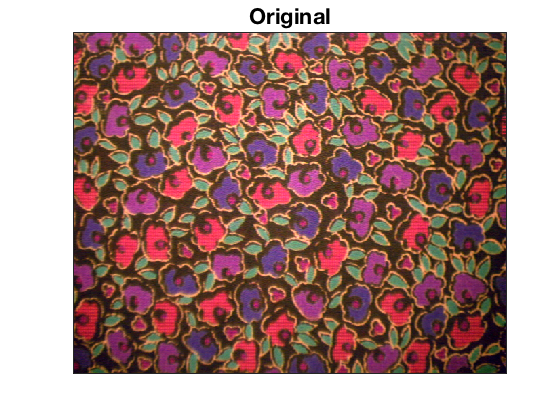}
        \caption{}
        \label{fig:original_texture}
    \end{subfigure}
    \begin{subfigure}{0.31\linewidth}
    \centering
        \includegraphics[width=\linewidth]{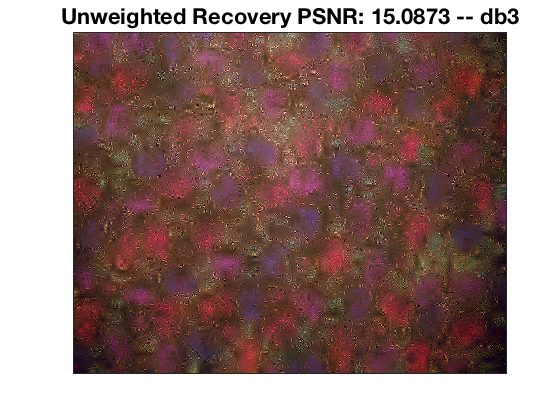}
        \caption{}
        \label{fig:unweighted_texture}
    \end{subfigure}
    \begin{subfigure}{0.31\linewidth}
        \includegraphics[width=\linewidth]{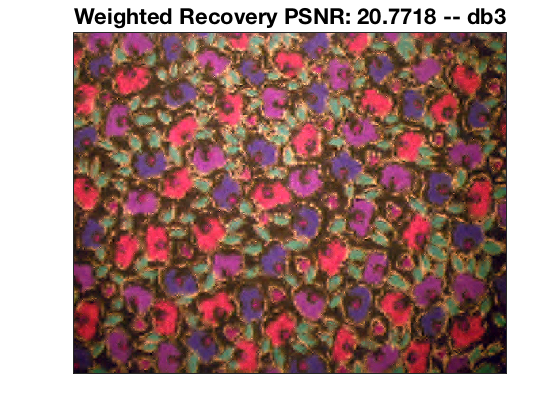}
        \caption{}
        \label{fig:weighted_texture}
    \end{subfigure}
    \caption{\eqref{fig:original_texture} A $480 \times 640$ pixel image of a wallpaper which has a repeating pattern. 
                  \eqref{fig:unweighted_texture} The recovery of the wallpaper image using unweighted $\ell_1$-minimization and 
                  the db3 basis from 10\% of the pixels.
                  \eqref{fig:weighted_texture} The recovery of the circles cartoon using weighted $\ell_1$-minimization and 
                  the db3 basis from 10\% of the pixels.
                  } 
    \label{fig:texture}
\end{figure*}

\begin{figure*} 
    \centering
    \begin{subfigure}{0.31\linewidth}
    \centering
        \includegraphics[width=\linewidth]{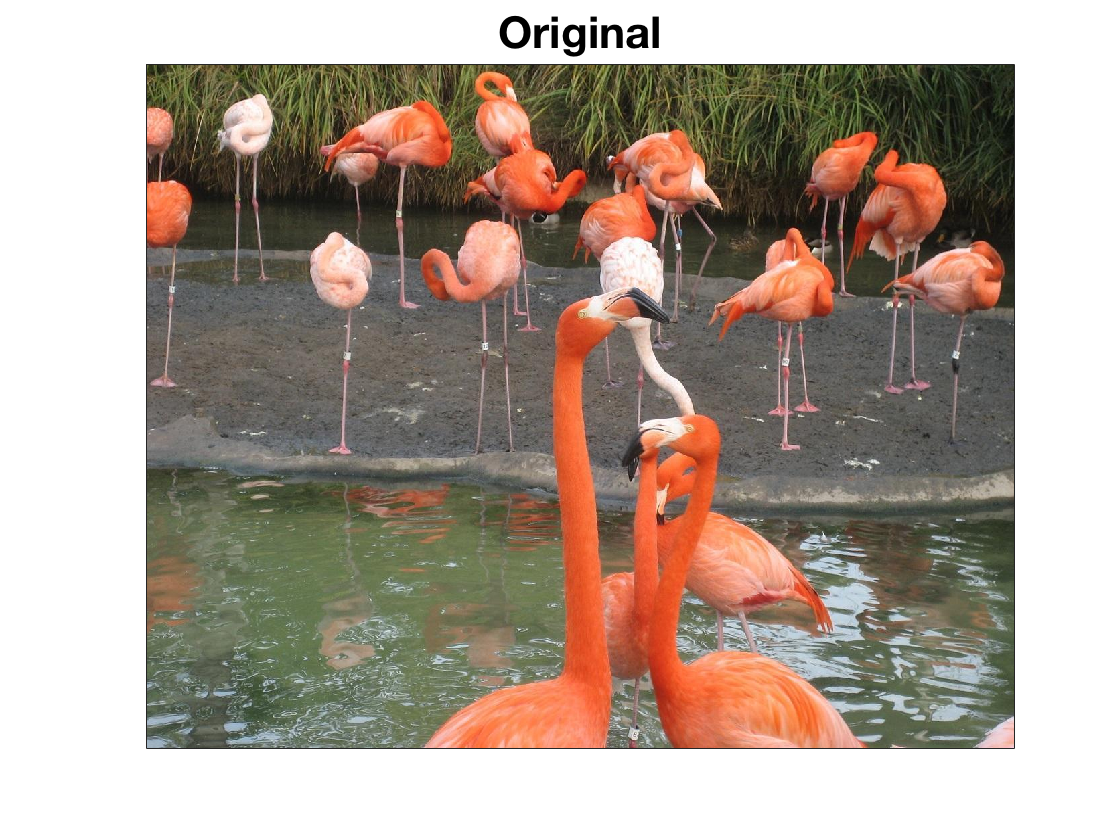}
        \caption{}
        \label{fig:original_flamingo}
    \end{subfigure}
    \begin{subfigure}{0.31\linewidth}
    \centering
        \includegraphics[width=\linewidth]{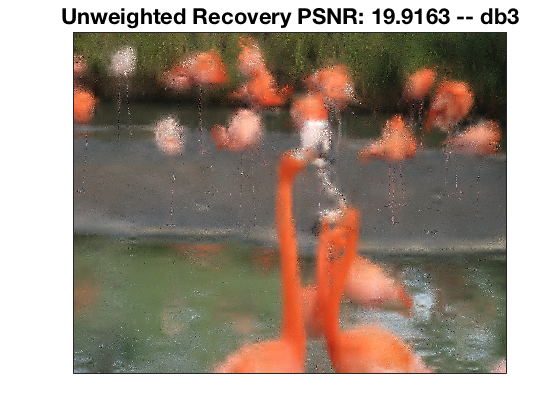}
        \caption{}
        \label{fig:unweighted_flamingo}
    \end{subfigure}
    \begin{subfigure}{0.31\linewidth}
        \includegraphics[width=\linewidth]{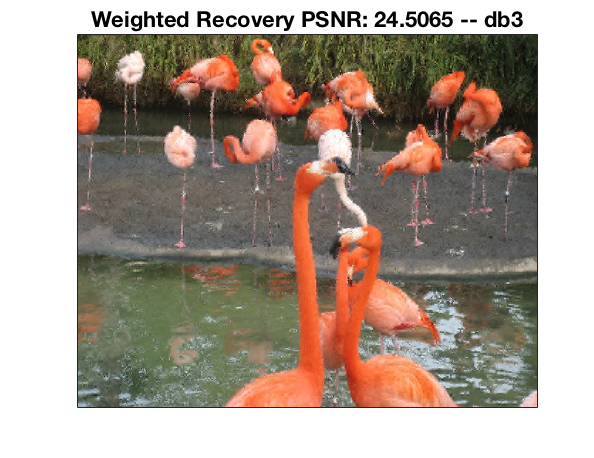}
        \caption{}
        \label{fig:weighted_flamingo}
    \end{subfigure}
    \caption{\eqref{fig:original_flamingo} A $972 \times 1296$ pixel image of flamingos which contains many 
                  different textures and shapes. 
                  \eqref{fig:unweighted_flamingo} The recovery of the flamingos using unweighted $\ell_1$-minimization and 
                  the db3 basis from 8\% of the pixels.
                  \eqref{fig:weighted_flamingo} The recovery of the flamingos using weighted $\ell_1$-minimization and 
                  the db3 basis from 8\% of the pixels.
            } 
    \label{fig:flamingo}
\end{figure*}

\begin{figure*} 
    \centering
    \begin{subfigure}{0.31\linewidth}
    \centering
        \includegraphics[width=\linewidth]{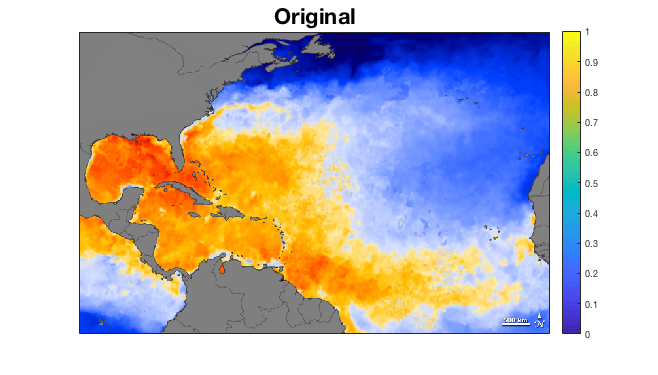}
        \caption{}
        \label{fig:original_ocean}
    \end{subfigure}
    \begin{subfigure}{0.31\linewidth}
    \centering
        \includegraphics[width=\linewidth]{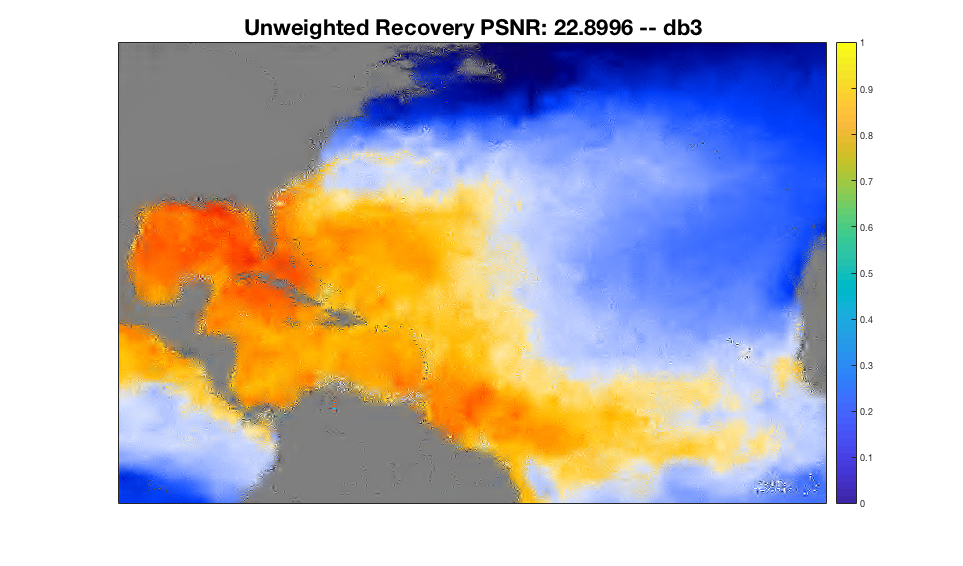}
        \caption{}
        \label{fig:unweighted_ocean}
    \end{subfigure}
    \begin{subfigure}{0.31\linewidth}
        \includegraphics[width=\linewidth]{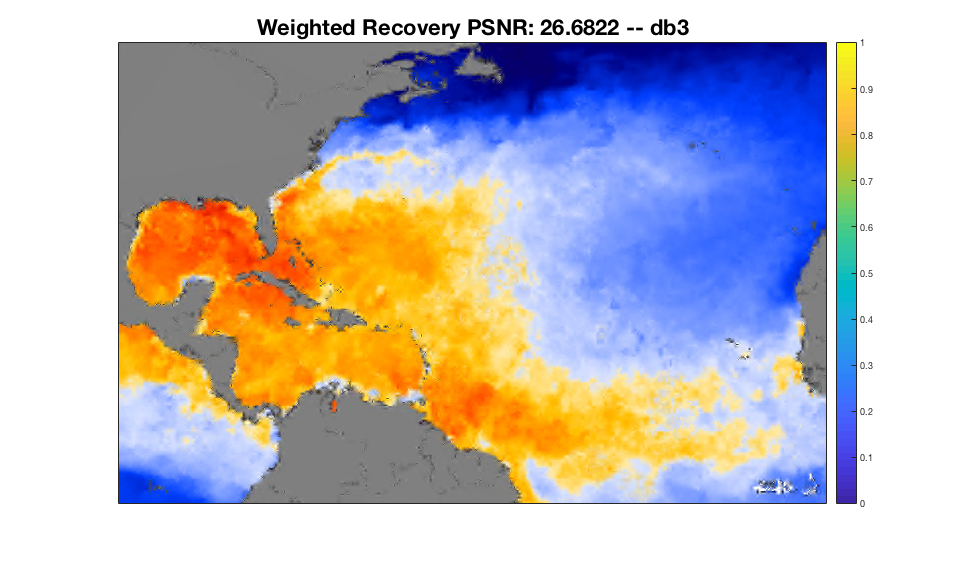}
        \caption{}
        \label{fig:weighted_ocean}
    \end{subfigure}
    \caption{\eqref{fig:original_ocean} A $480 \times 720$ pixel plot of the surface temperature of the Atlantic oceanns many different textures and shapes. 
                  \eqref{fig:unweighted_ocean} The recovery of the ocean temperatures using unweighted $\ell_1$-minimization and 
                  the db3 basis from 14\% of the pixels.
                  \eqref{fig:weighted_ocean} The recovery of the temeratures using weighted $\ell_1$-minimization and 
                  the db3 basis from 14\% of the pixels.
                  } 
    \label{fig:science}
\end{figure*}

Here we consider more color examples for a variety of image types, namely, cartoons in Figure \ref{fig:cartoon},
textures in Figure \ref{fig:texture}, natural scenes with animals in Figure \ref{fig:flamingo},
and images genereated from scientific data in Figure \ref{fig:science}.

\bibliography{references}{}
\bibliographystyle{plain}

\end{document}